\documentclass[twocolumn,epjc3]{svjour3}  
\usepackage{amsmath,amssymb,amsfonts}
\usepackage{graphicx,graphics}
\usepackage{graphics}
\usepackage{url}
\usepackage{slashed}
\usepackage{xcolor}
\usepackage[normalem]{ulem}
\usepackage{mathtools}

\journalname{Eur. Phys. J. C}

\def\gapproxeq{\lower .7ex\hbox{$\;\stackrel{\textstyle                                                    
>}{\sim}\;$}}  

\newcommand \Pomeron {I\!\!P}

\begin{document}

\title{Predictions for exclusive $\Upsilon$ photoproduction in ultraperipheral ${\rm Pb}+{\rm Pb}$ collisions at the LHC at next-to-leading order in perturbative QCD}

\author{Kari~J.~Eskola$^{1,2,}$\thanksref{e2},\,C.~A.~Flett$^{1,2,3,}$\thanksref{e3},\,V.~Guzey$^{1,2,}$\thanksref{e1},\,T.~L\"oyt\"ainen$^{1,2,}$\thanksref{e4},\,H.~Paukkunen$^{1,2,}$\thanksref{e5}}

\thankstext{e2}{e-mail: kari.eskola@jyu.fi}
\thankstext{e3}{e-mail: christopher.flett@ijclab.in2p3.fr}
\thankstext{e1}{e-mail: vadim.a.guzey@jyu.fi}
\thankstext{e4}{e-mail: topi.m.o.loytainen@jyu.fi}
\thankstext{e5}{e-mail: hannu.paukkunen@jyu.fi}

\institute{University of Jyvaskyla, Department of Physics, P.O. Box 35, FI-40014 University of Jyvaskyla, Finland \and
Helsinki Institute of Physics, P.O. Box 64, FI-00014 University of Helsinki, Finland \and
Universit\'{e} Paris-Saclay, CNRS, IJCLab, 91405 Orsay, France
}

\maketitle

\abstract{
We present predictions for the rapidity-\linebreak differential cross sections of exclusive $\Upsilon$ photoproduction in ultraperipheral collisions (UPCs) of lead ions at the Large Hadron Collider (LHC). We work in the framework of collinear factorization at next-to-leading order (NLO) in perturbative QCD, modeling the generalized parton distributions (GPDs) through the Shuvaev transform of nuclear parton distribution functions (nPDFs). While the effects due to the GPD modeling turn out to be small, the direct NLO predictions still carry significant nPDF-originating uncertainties and depend strongly on the choices of the factorization and renormalization scales. To tame the scale dependence and to account for the fact that the NLO calculations generally underpredict the photoproduction measurements on protons, we also present alternative, data-driven predictions. In this approach the underlying photoproduction cross sections on lead are found by combining their nuclear modifications calculated at NLO with the measured photoproduction cross sections on protons. The data-driven strategy reduces the uncertainties associated with the scale choices, and essentially eliminates the effects of GPD modeling thereby leaving the cross sections sensitive mainly to the input nPDFs. Our estimates indicate that the process is measurable in ${\rm Pb}+{\rm Pb}$ collisions at the LHC. 
}

\section{Introduction}
\label{Sec:Intro}

The exclusive production of heavy vector mesons $V$ in ultraperipheral collisions (UPCs) of heavy nuclei, $A_1 + A_2 \rightarrow A_1 + V + A_2$, has for a long time captured the interest of both the theoretical and experimental high-energy physics communities. It allows one to study not only the perturbative aspects of Quantum Chromodynamics (QCD) but to also probe the non-perturbative structure of nuclei \cite{Bertulani:2005ru,Baltz:2007kq,Contreras:2015dqa,Klein:2019qfb}. These ultraperipheral events are largely initiated by electromagnetic interactions as the short-range hadronic interactions are strongly suppressed by the exclusivity of the final state. The vector meson production then effectively proceeds through an interaction of a quasi-real photon from one nucleus with the other nucleus such that the colliding nuclei remain intact and the exclusivity of the vector meson production is maintained via a net-colourless production mechanism. At leading order (LO) in perturbative QCD (pQCD) \cite{Ryskin:1992ui}, the production is mediated through a two-gluon exchange, while at next-to-leading order (NLO), there is also a quark-pair initiated contribution~\cite{Ivanov:2004vd}. The exchanged partons carry different longitudinal momentum fractions depending on an additional off-forward skewness parameter, $\xi$. This results in a factorization~\cite{Collins:1996fb} of the scattering amplitude into the perturbatively calculable hard-scattering part and non-perturbative generalized parton distribution functions (GPDs)~\cite{Ji:1996nm,Radyushkin:1997ki,Diehl:2003ny}. 

The first UPC measurements of exclusive $J/\psi$ mesons came from the PHENIX collaboration at the Relativistic Heavy Ion Collider (RHIC) in Au+Au collisions at the nucleon-nucleon centre-of-mass system (c.m.s.) energy of $\sqrt{s_{NN}} = 200$ GeV~\cite{PHENIX:2009xtn}. Subsequently, the ALICE, CMS, and LHCb collaborations at the Large Hadron Collider (LHC) have measured the same process in heavier ${\rm Pb}+{\rm Pb}$ UPCs at $\sqrt{s_{NN}}$ = 2.76 and 5.02 TeV in a wide range of the $J/\psi$ rapidities from $y=0$ up to $|y| \sim 4.5$~\cite{ALICE:2012yye,ALICE:2013wjo,CMS:2016itn,ALICE:2019tqa,ALICE:2021gpt,LHCb:2021bfl,LHCb:2022ahs}. These data -- not forgetting the multitude of statistics anticipated in the heavy-ion programme of the High Luminosity LHC \cite{Citron:2018lsq} -- provide ample grounds for understanding the perturbative structure of QCD and the nuclear shadowing phenomenon encoded e.g. in nuclear parton distribution functions (nPDFs) \cite{Kusina:2020lyz,Eskola:2021nhw,Helenius:2021tof,AbdulKhalek:2022fyi}, down to momentum fractions of $x \sim (M_V/\sqrt{s_{NN}}) \exp(-|y|) \sim 10^{-5}$ at resolution scales $\mu^2 \sim {\cal O}(M_V^2)$, where $M_V$ is the mass of the vector meson. 

In our previous works~\cite{Eskola:2022vpi,Eskola:2022vaf}, we studied the exclusive photoproduction of $J/\psi$ mesons in ${\rm Pb}+{\rm Pb}$ and ${\rm O}+{\rm O}$ collisions to next-to-leading order (NLO) in pQCD. By approximating the GPDs with PDFs, we demonstrated the complicated interplay of the quark and gluon contributions at NLO over the entire LHC acceptance in rapidity, and showed that our theoretical predictions agree with the experimental data for this process~\cite{ALICE:2012yye,ALICE:2013wjo,CMS:2016itn,ALICE:2019tqa,ALICE:2021gpt,LHCb:2021bfl,LHCb:2022ahs} within the large theoretical uncertainties associated with the choice of the factorization/renormalization scales and nPDFs. Valuable and complementary information on nPDFs at small momentum fractions $x$, in particular, on the scale dependence of nuclear shadowing, can be obtained by  studying exclusive photoproduction of heavier vector mesons such as $\Upsilon$ mesons consisting of a bottom quark and its antiquark. To date, while there have been measurements of the exclusive photoproduction of $\Upsilon$ in $e+p$ collisions at Hadron Electron Ring Accelerator (HERA) \cite{H1:2000kis,ZEUS:1998cdr,ZEUS:2009asc}, as well as in $p+p$~\cite{LHCb:2015wlx} and $p+{\rm Pb}$ collisions at the LHC \cite{CMS:2018bbk}, there has been no reported measurement of the exclusive production of $\Upsilon$ mesons in heavy-ion collisions. 

In the work presented here, we make predictions for the rapidity-differential cross sections of this process at $\sqrt{s_{NN}} = 5.02\,{\rm TeV}$ in ${\rm Pb}+{\rm Pb}$ collisions, extending our previous framework to incorporate a more careful GPD modeling by relating the nPDF to nuclear GPDs through the so-called Shuvaev integral transform~\cite{Shuvaev:1999fm,Shuvaev:1999ce,Golec-Biernat:1999trj}. Despite the larger interaction scale in comparison to the $J/\psi$ case, the theoretical uncertainties in the case of $\Upsilon$ production are still sizable and -- as was already noticed in the pioneering work of Ref.~\cite{Ivanov:2004vd} and as we confirm in this work as well -- natural choices of the factorization/renormalization scales $\mu^2 \sim {\cal O}(M_\Upsilon^2)$ do not lead to a particularly good description of the HERA data. This would then cast doubts also on our direct NLO predictions in ${\rm Pb}+{\rm Pb}$. As a workaround, we will adopt an alternative method in which we anchor our predictions for the underlying $\gamma+{\rm Pb} \to \Upsilon+{\rm Pb}$ cross sections on the HERA data on the $\gamma+p \to \Upsilon+p$ process by using the NLO calculations only for the ratios of cross sections between these two processes. We call this method the data-driven approach. We also analyze the nuclear modifications of the $\gamma+{\rm Pb} \to \Upsilon+{\rm Pb}$ cross sections due to nuclear effects in PDFs and show that for $\xi < 10^{-3}$, they coincide very closely with the gluon nuclear modification factor squared. 

The rest of the paper is organised as follows. In Sec.~\ref{Sec:2A}, we summarize our theoretical framework for the exclusive photoproduction of $\Upsilon$ in ultraperipheral ${\rm Pb}+{\rm Pb}$ collisions within NLO pQCD, and then discuss the modeling of GPDs in Sec.~\ref{sec:Shuvaev}. The ingredients of our data-driven approach are explained in Sec.~\ref{sec:Datadriven}. In Sec.~\ref{sec:Results}, we then present our results for the cross sections and their nuclear modifications, discussing also how our calculations build up from various components. Finally, we draw our conclusions in Sec.~\ref{sec:Conclusions}, outlining also future directions.

\section{Theoretical framework}
\label{Sec:2}

\subsection{Exclusive $\Upsilon$ photoproduction in ${\rm Pb}+{\rm Pb}$ UPCs at NLO pQCD}
\label{Sec:2A}

Within the equivalent-photon approximation~\cite{Bertulani:2005ru,Baltz:2007kq}, the rapidity-differential cross section for the process ${\rm Pb} + {\rm Pb} \rightarrow {\rm Pb} + \Upsilon + {\rm Pb}$  can be written as
\begin{align}
\label{XS_plus_minus}
    \frac{\text{d} \sigma^{ {\rm Pb} + {\rm Pb} \rightarrow {\rm Pb} + \Upsilon + {\rm Pb}} }{\text{d} y}  & =  \left(k \frac{dN_{\gamma}^{\rm Pb} (k)}{dk} \sigma^{\gamma(k) {\rm Pb} \rightarrow \Upsilon {\rm Pb}}\right)_{k=k^+}  \\\
    & + \left(k \frac{dN_{\gamma}^{\rm Pb} (k)}{dk} \sigma^{{\rm Pb} \gamma(k) \rightarrow {\rm Pb}  \Upsilon}\right)_{k=k^-} \,, \nonumber
\end{align}
where $kdN_{\gamma}^{{\rm Pb}} (k)/dk$ is the Weizs\"{a}cker-Williams~(WW) number density or flux of photons from the Pb nucleus as a function of the photon energy $k^{\pm} = (M_{\Upsilon}/2) \exp(\pm y)$ with $M_{\Upsilon}$ being the mass of the $\Upsilon$ meson. The cross sections for the underlying photoproduction subprocesses are labelled by $\sigma^{{\rm Pb} \gamma(k^-)  \rightarrow {\rm Pb}  \Upsilon}$ and $\sigma^{\gamma(k^+) {\rm Pb} \rightarrow \Upsilon {\rm Pb}}$. The two terms in Eq.~(\ref{XS_plus_minus}) correspond to the right-moving and left-moving photon sources, which results in a two-fold ambiguity of the photon energy at a given value of $y \neq 0$.

The WW flux is given by a convolution of the impact-parameter dependent photon flux $N_{\gamma}^A(k,\vec{b})$ calculable in QED~\cite{Vidovic:1992ik} and the nuclear suppression factor $\Gamma_{AA}(\vec{b})$,
\begin{equation}
k \frac{dN_{\gamma }^A (k)}{dk} = \int d^2 \Vec{b} \, N_{\gamma }^A (k,\Vec{b}) \Gamma_{AA}(\Vec{b}) \,. 
\label{eq:flux1} 
\end{equation}
Here, $\vec b$ is the two-dimensional vector between the centres of the two colliding Pb nuclei in the transverse plane, and $\Gamma_{AA}(\vec{b})$ encodes the Glauber-model probability of having no additional hadronic interaction in the event;  for details see~\cite{Eskola:2022vpi}.

The cross section for the photoproduction process mediating the ultraperipheral ${\rm Pb} + {\rm Pb} \rightarrow {\rm Pb} + \Upsilon + {\rm Pb }$ reaction can be expressed in terms of the exclusive photoproduction cross section per bound nucleon $N$, $ d\sigma_A^{\gamma N \rightarrow \Upsilon N}(W)/dt$, and the nuclear form factor $F_A(t)$ as 
\begin{equation}
\sigma^{\gamma A \rightarrow \Upsilon A}(W) = \frac{ \text{d} \sigma_A^{\gamma N \rightarrow \Upsilon N}(W)}{\text{d} t}\biggr|_{t=0}\int_{|t_{\rm min}|}^{\infty} \text{d}t |F_A(-t)|^2 \,,
\label{eq:cs1}
\end{equation}
where 
\begin{equation}
\frac{ \text{d} \sigma_A^{\gamma N \rightarrow \Upsilon N}(W)}{ \text{d} t}\biggr|_{t=0} = \frac{|\mathcal M_A^{\gamma N \rightarrow \Upsilon N}|^2}{16 \pi W^4}
\label{eq:cs2}
\end{equation}
is the $t$-differential cross section evaluated at $t=0$, the variable $t$ being the squared momentum transfer in the process, $W$ is the $\gamma{\text -}N$ c.m.s.~energy, and
$|t_{\rm min}|=m_N^2 (M_{\Upsilon}^2/W^2)^2$ is the minimal momentum transfer \linebreak  squared with $m_N$ denoting the nucleon mass.

The nuclear form factor $F_A(t)$ is well known from measurements of elastic electron-nucleus scattering and for heavy nuclei it is typically given by the Fourier transform of the two-parameter Woods-Saxon charge distribution $\rho(r)$~\cite{Woods:1954zz},
\begin{equation}
F_A(t) = \int d^3 r\,e^{i {\mathbf q} \cdot {\mathbf r}} \rho(r)  \,,
\label{eq:F_A}
\end{equation}
where
\begin{equation}
\rho(r) = \frac{\rho_0}{1+\exp \left(\frac{r-R_A}{d}\right)} \,,
\end{equation}
with $|{\mathbf q}| = \sqrt{-t}$. We take $d=0.546$ fm for the nucleus skin depth and $R_A /{\rm fm} = 1.12 A^{1/3} - 0.86 A^{-1/3}$ for the nuclear radius. The normalization $\rho_0 \approx 0.17~{\rm fm}^{-3}$ is fixed by requiring that $F_A(0) =A= 208$ for Pb. 

The hard scattering amplitude for exclusive $\Upsilon$ photoproduction per nucleon $N$ bound in the nucleus $A$ can be described at NLO in collinear factorization by~\cite{Ivanov:2004vd},
\begin{align}
\mathcal M_A^{\gamma N \rightarrow \Upsilon N}(\xi, t=0) & = \frac{ 4 \pi \sqrt{4 \pi \alpha}  e_b (\epsilon^{\ast}_\Upsilon \cdot \epsilon_\gamma) }{ N_c } \left (  \frac{ \langle O_1 \rangle_\Upsilon }{m_b^3 } \right)^{1/2}  \nonumber \\
& \times I(\xi, t=0) \,,
\label{eq:M}
\end{align}
where 
\begin{align}
\label{conv}
I(\xi,t=0) & = \int_{-1}^1  \text{d} x \bigg[ T_g(x,\xi,\mu_R,\mu_F) F^g(x,\xi,t=0,\mu_F)  \nonumber \\
& + T_q(x,\xi,\mu_R,\mu_F)F^{q,S}(x,\xi,t=0,\mu_F)\bigg] \,.
\end{align}
In Eq.~(\ref{eq:M}), $e_b=1/3$ and $m_b=M_{\Upsilon}/2$ are the electric charge and the mass of the bottom quark, respectively;
$\alpha$ is the fine-structure constant; $N_c=3$ is the number of colors; $\epsilon_{\gamma}$ and $\epsilon_{\Upsilon}^{\ast}$ are the polarization 
vectors of the initial-state photon and the final-state vector meson, respectively; 
$\langle O_1 \rangle_\Upsilon$ is the non-relativistic QCD (NRQCD) matrix element for the $\Upsilon \to b \bar{b}$ transition, which
is proportional to the radial $\Upsilon$ wavefunction at the origin and which is fixed by the experimental value of the $\Upsilon$ decay width to a dilepton pair, see~\cite{Hoodbhoy:1996zg}. Note that in this approach, $M_{\Upsilon}=2 m_b$.

The reduced matrix element $I(\xi,t=0)$ is given by a convolution of the gluon $T_g(x,\xi,\mu_R,\mu_F)$ and quark $T_q(x,\xi,\mu_R,\mu_F)$ NLO coefficient functions with the  
gluon $F^g(x,\xi,t,\mu_F)$ and quark singlet $F^{q,S}(x,\xi,t,\mu_F)$ matrix elements involving the corresponding GPDs.
Note that the coefficient functions depend on the longitudinal momentum fraction $x$, the skewness $\xi=M_{\Upsilon}^2/(2 W^2-M_{\Upsilon}^2)$, the renormalization scale $\mu_R$, and the factorization scale $\mu_F$. In our analysis, we set $\mu=\mu_R=\mu_F$ and vary $\mu$ in the $m_b/2 \leq \mu \leq 2 m_b$ interval.

In the leading-twist approximation and neglecting the mass of the nucleons, the factors $F^g$ and $F^{q,S}$ in the $t=0$ limit can be expressed in terms
of the helicity-conserving gluon $H^g(x,\xi,t,\mu_F)$ and quark singlet \linebreak $H^{q,S}(x,\xi,t,\mu_F)$ GPDs as follows~\cite{Diehl:2003ny},
\begin{align}
F^g(x, \xi, t = 0, \mu_F) & = \sqrt{1-\xi^2}H^g(x, \xi, t = 0, \mu_F), \\
F^{q,S}(x, \xi, t = 0, \mu_F) & = \sqrt{1-\xi^2}H^{q,S}(x, \xi, t = 0, \mu_F) \,, \nonumber
\end{align}
with
\begin{align}
H^{q,S}(x, \xi, t = 0, \mu_F) & = \sum_{
q=u,d,s,c
} 
\Big[
H^{q}(x, \xi, t = 0, \mu_F) \nonumber \\ & - H^{q}(-x, \xi, t = 0, \mu_F) \Big] \,. 
\end{align}
At $\xi=t=0$, these GPDs reduce to the usual gluon, quark, and antiquark PDFs of the (bound) nucleons,
\begin{align}
H^g(\pm x, \xi = 0, t = 0, \mu_F) & =  xg(x,\mu_F) \,, \nonumber\\
H^q(x, \xi = 0, t = 0, \mu_F)     & = q(x, \mu_F) \,, \nonumber\\
H^q(-x, \xi = 0, t = 0, \mu_F)    & =  -\bar{q}(x,\mu_F) \\ 
 H^{q,S}(x, \xi = 0, t = 0, \mu_F)  & = \sum_{
 q=u,d,s,c
 } \nonumber
 \Big[ q(x, \mu_F) + \bar{q}(x,\mu_F) \Big] \\ & \equiv q^S(x,\mu_F)  \nonumber \,. 
\end{align}
where $x \in [0,1]$.

In Eq.~\eqref{conv}, each value of the skewness parameter $\xi$ entails an integration over the convolution variable $x$. 
In the literature, see~\cite{Diehl:2003ny} for review, the $|x| \geq \xi$ interval is called the DGLAP region since GPDs there can be interpreted as parton distribution functions evolving in $\log (\mu_F^2)$ according to the modified Dokshitzer-Gribov-Lipatov-Altarelli-Parisi (DGLAP) evolution equations. The $|x| < \xi$ interval is called the ERBL region because GPDs there resemble parton distribution amplitudes, whose $\mu_F$ evolution is given by
the modified Efremov-Radyushkin-Brodsky-Lepage (ERBL) evolution equations. In this work, we employ the Shuvaev transform at NLO to model the $\xi$ dependence of GPDs and to relate the GPDs to PDFs in the DGLAP region, see details in Sec.~\ref{sec:Shuvaev}. To counteract the possible invalidity of the Shuvaev transform in the time-like ERBL region of $|x| < \xi$, we convolute the GPDs with only the imaginary part of the gluon and quark coefficient functions in Eq.~\eqref{conv}, which vanish identically for $|x| < \xi$. We then restore the real part via the high-energy dispersion 
relation~\cite{Ryskin:1995hz}
\begin{align}
\label{re}
& \frac{\Re e {\mathcal M}_A^{\gamma N \rightarrow \Upsilon N}(\xi,t=0)}{\Im m {\mathcal M}_A^{\gamma N \rightarrow \Upsilon N}(\xi,t=0)} \\
& ={\rm tan}\left(\frac{\pi}{2}~\frac{\partial \ln (\Im m \mathcal M_A^{\gamma N \rightarrow \Upsilon N}(\xi,t=0)/(1/\xi))}{\partial \ln (1/\xi) }\right) \,. \nonumber
\end{align}
We have checked that this relation accurately reproduces the directly computed real part contribution for $W \gapproxeq 40$~GeV at a percent level in the case that GPDs are approximated by PDFs. At smaller $W$ the deviation increases but, as will be discussed, our main data-driven predictions will nevertheless be valid only for $W \gtrsim 100 \, {\rm GeV}$.

To summarize, the standard pQCD approach to the calculation of the $\sigma^{\gamma A \to \Upsilon A}(W)$ cross section is based on 
Eqs.~(\ref{eq:cs1}) and (\ref{eq:cs2}), where the hard scattering nuclear amplitude per bound nucleon  
$\mathcal M_A^{\gamma N \rightarrow \Upsilon N}(\xi, t=0)$ is calculated using the bound nucleon (nucleus) gluon and quark GPDs, 
see Eqs.~(\ref{eq:M}) and (\ref{conv}). Replacing the bound nucleon by the free proton in these equations, one readily obtains the NLO pQCD predictions for the proton target. The cross section of exclusive $\Upsilon$ photoproduction on the proton reads [compare to Eq.~(\ref{eq:cs1})]
\begin{equation}
\sigma^{\gamma p \rightarrow \Upsilon p}(W) = \frac{1}{B_{\Upsilon}(W)}  \frac{ \text{d} \sigma^{\gamma p \rightarrow \Upsilon p}(W) }{ \text{d} t} \biggr|_{t=0}, 
\label{eq:cs1_p}
\end{equation}
where $B_{\Upsilon}(W)$ is the energy-dependent slope of the $t$ dependence of the $\gamma+ p \to \Upsilon+p$ cross section, which is assumed to be 
exponential;
$\text{d} \sigma^{\gamma p \rightarrow \Upsilon p}(W)/ \text{d} t (t=0)$ is the differential cross section at $t=0$, which is
calculated using Eqs.~(\ref{eq:cs2}), (\ref{eq:M}) and (\ref{conv}) with nuclear GPDs replaced by their free-proton counterparts.

The $t$ dependence of the $\gamma+ p \to \Upsilon+p$ cross section has never been measured. Therefore, for the $B_{\Upsilon}(W)$ slope, we use the following parametrization motivated by Regge phenomenology, 
\begin{equation}
B_{\Upsilon}(W)  = B_0 + 4 \alpha_{\Pomeron}^{\prime} \ln \left(\frac{W}{W_0}\right) \,,
\label{eq:B_slope}
\end{equation}
where $B_0 = 4.63$ GeV$^{-2}$, $\alpha_{\Pomeron}^{\prime} = 0.06$ GeV$^{-2}$, and $W_0 = 90$ GeV.
While the value of $B_0$ is compatible with fits to the $t$ dependence of elastic $J/\psi$ photoproduction on the proton 
at HERA~\cite{H1:2000kis,H1:2013okq}, the value of slope of the Pomeron trajectory $\alpha_{\Pomeron}^{\prime}$ is fixed by
Model 4 of~\cite{Khoze:2013dha}, which fits a wide variety of data on diffraction in proton-proton scattering at the LHC.

\subsection{GPD modeling}
\label{sec:Shuvaev}

Generalized parton distributions naturally appear in the framework of collinear factorization for hard exclusive processes~\cite{Collins:1996fb} and combine properties of usual PDFs, distribution amplitudes and elastic form factors \cite{Ji:1996nm,Radyushkin:1997ki,Diehl:2003ny}. Since GPDs depend on two light-cone momentum fractions $x$ and $\xi$, the invariant momentum transfer squared $t$, and the factorization scale $\mu_F$, their modeling and extraction from the available experimental data has been notoriously challenging, see, e.g.~\cite{Berthou:2015oaw}. However, at small values of the skewness $\xi$, GPDs rather closely resemble usual PDFs in the $|x| \geq \xi$ DGLAP region and the $|x| < \xi$ ERBL region plays typically only a minor role. These facts significantly simplify the modeling of GPD-originating effects even if the experimental constraints for the three-dimensional structure of GPDs are weak. 

One of the most widely used models of GPDs at small $\xi$ is based on the so-called Shuvaev transform, which is a method to analytically solve the LO $Q^2$ evolution equations of GPDs~\cite{Shuvaev:1999fm,Shuvaev:1999ce,Golec-Biernat:1999trj}. It is a generalization of solving the usual DGLAP evolution equations using Mellin moments of PDFs. To briefly summarize the method, one first defines effective PDFs, whose Mellin moments are equal to the Gegenbauer (conformal) moments of GPDs. One then inverts these relations and expresses GPDs as certain integrals of the effective PDFs at any given factorization scale $\mu_F$. Finally, using the condition of polynomiality of the conformal moments (see details in~\cite{Shuvaev:1999ce,Martin:2008gqx}), one argues that the effective PDFs can be approximated by the usual PDFs and obtains the desired connection between GPDs at small $\xi$ and PDFs. In other words, the input GPDs at some low scale $\mu_0$ are assumed to be independent of $\xi$, and the $\xi$ dependence is then generated radiatively during the scale evolution -- this warrants to speak about perturbative skewness. Moreover, since the mixing of the conformal moments under the NLO $Q^2$ evolution is suppressed by powers of $\xi$, the Shuvaev transform can also be safely used at NLO in the $\xi \ll 1$ limit~\cite{Shuvaev:1999fm}. As a phenomenological application of the method, it was shown in NLO and next-to-next-to-leading order (NNLO) analyses~\cite{Kumericki:2009uq} that a flexible parametrization of quark and gluon GPDs of the proton in terms of their conformal moments describes well the available HERA data on deeply virtual Compton scattering (DVCS) on the proton. In the case that the condition $\xi \ll 1$ is not met, the Shuvaev transform should be substituted by explicitly solving the GPD evolution equations \cite{Bertone:2022frx,Dutrieux:2023qnz}.

In our work, we employ the Shuvaev transform at NLO as a means to relate the GPDs to PDFs in the DGLAP region. Thus, 
the quark and gluon GPDs are obtained as integrals of the corresponding quark and gluon PDFs,
\begin{align}
& H^q(x,\xi,t=0,\mu_F)  =  \nonumber \\ 
& \int_{-1}^1\text{d}x^{\prime}~ \biggl[ \frac{2}{\pi} \Im m \int_0^1 \frac{\text{d}s}{y(s) \sqrt{1-y(s)x^{\prime}}} \biggr] \frac{\text{d}}{\text{d}x^{\prime}} \frac{q(x^{\prime},\mu_F)}{|x^{\prime}|}\,, \nonumber\\
& H^g(x, \xi,t=0, \mu_F) =  \label{Shuvaevt} \\
& \int_{-1}^1\text{d}x^{\prime}~ \biggl[ \frac{2}{\pi} \Im m \int_0^1 \frac{\text{d}s~(x+\xi(1-2s)}{y(s) \sqrt{1-y(s)x^{\prime}}} \biggr] \frac{\text{d}}{\text{d}x^{\prime}} \frac{g(x^{\prime},\mu_F)}{|x^{\prime}|} \,,
\nonumber
\end{align}
where the kernel of the transform is
\begin{equation} y(s) = \frac{4s(1-s)}{x+\xi(1-2s)} \,. 
\label{ys} \end{equation} 
As we explained above, Eq.~(\ref{Shuvaevt}) is used only to calculate the imaginary part of the hard scattering amplitude ${\mathcal M}_A^{\gamma N \rightarrow \Upsilon N}(\xi,t=0)$. The real part probing the ERBL region is restored via the high-energy dispersion relation~(\ref{re}). In practice, the Shuvaev integrals in Eqs.~(\ref{Shuvaevt}) involving derivatives of the input PDFs converge rather slowly and have to be precomputed before evaluating Eq.~(\ref{conv}). To this end, we have computed the GPDs in a three-dimensional $x, \xi/x, \mu^2$ grid using Eqs.~(\ref{Shuvaevt}). The construction of the GPD grid is optimised such that areas in the parameter space that result in a flat interpolation are not overly populated: having more points around $\xi/x \sim 1$ mitigates edge effects at the boundary of the DGLAP and ERBL regions~\cite{Martin:2008gqx}, while the interpolation in $\mu^2$ is relatively smooth and requires fewer points. 

\begin{figure*}
\centering
\includegraphics[width=0.75\textwidth]{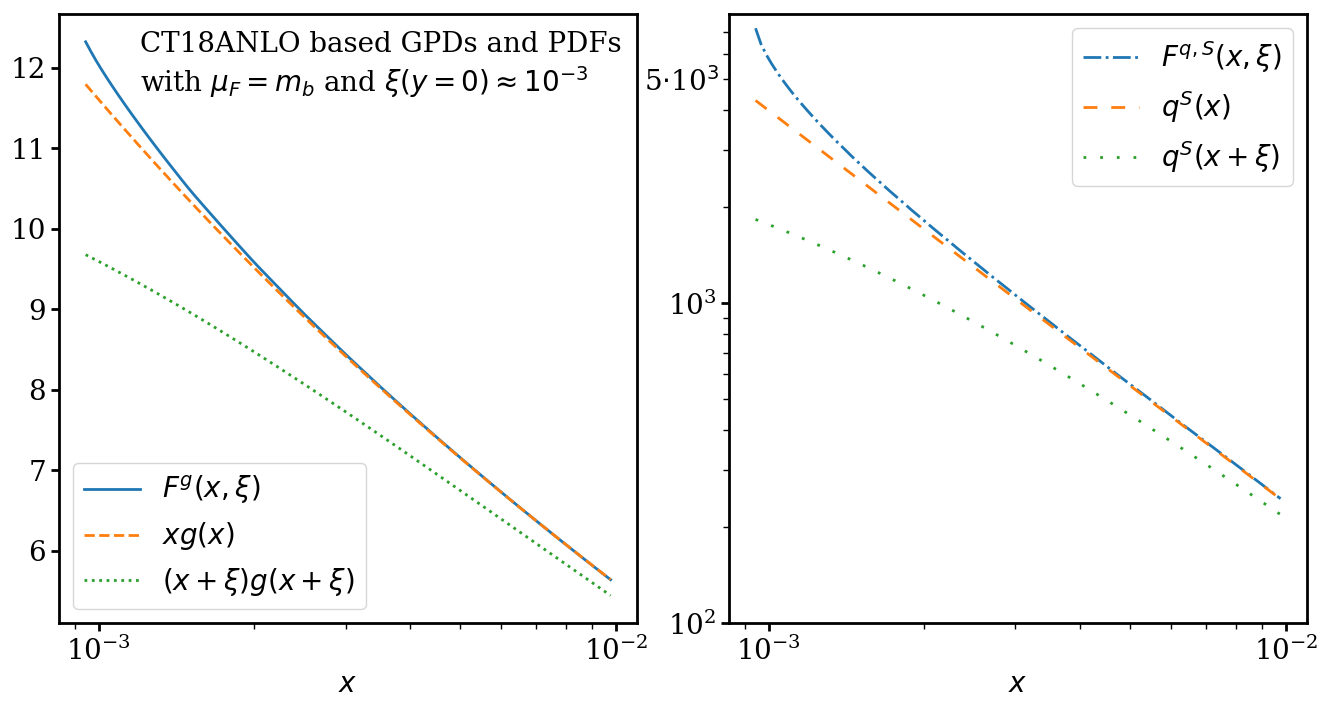}
\caption{
The gluon (left panel) and quark singlet (right panel) GPDs (blue curves) $F^g(x,\xi)$ and $F^{q,S}(x,\xi)$ with $\xi \approx 10^{-3}$ obtained through the Shuvaev transformation, compared with PDFs $xg(x)$ and $q^S(x)$ (orange dashed curves) at $\mu_F = m_b$ as a function of $x$. In addition, we also present the distributions $(x+\xi)g(x+\xi)$ and $q^S(x+\xi)$ (green dotted curves).
} 
\label{fig:gpds}
\end{figure*}

In Fig.~\ref{fig:gpds}, we illustrate the effect of finite skewness in GPDs by comparing the gluon and quark-singlet GPDs, $F^g(x,\xi)$ and $F^{q,S}(x,\xi)$, obtained through the Shuvaev transform, with their values at $\xi=0$, $xg(x)$ and $q^S(x)$, as a function of $x$ at the scale $\mu_F = m_b$. We have used here the CT18ANLO proton PDFs \cite{Hou:2019efy} taken from the LHAPDF library \cite{Buckley:2014ana}. In these plots, we have fixed $\xi \approx 10^{-3}$, which corresponds to the kinematic value of the skewness parameter probed in $\Upsilon$ photoproduction in ${\rm Pb}+{\rm Pb}$ UPCs at $5.02\,{\rm TeV}$ and $y=0$.  The distributions are plotted in a small interval of the DGLAP region, $x \in [\xi, 10^{-2}]$, where the Shuvaev transform is a reliable way to obtain the perturbatively generated skewness of the GPDs. One can see from the figure that the effect of skewness -- the deviation between the blue and orange curves -- is rather small for most values of $x$, but grows towards the point $x=\xi$, especially in the case of quarks. At the same time, to compare with the commonly used skewness factor due to the Shuvaev transform~\cite{Shuvaev:1999ce,Diehl:2007zu}, we also show the gluon and quark singlet PDFs evaluated at the $x+\xi$ point,  $(x+\xi)g(x+\xi,\mu_F)$ and $ q^S(x+\xi,\mu_F)$. In this case, the effect of skewness is noticeable  (the deviation between the blue and green lines is significant). However, in our NLO pQCD analysis the $(x+\xi)g(x+\xi,\mu_F)$ and $ q^S(x+\xi,\mu_F)$ PDFs do not play any special role and we find that the numerical effect of the skewness effects induced by the Shuvaev transform in the calculated cross sections of $\Upsilon$ photoproduction on the proton and a heavy nucleus is small. 

The effect of the Shuvaev transform is larger at larger scales $\mu_F$, where the effective power growth of the partons becomes steeper and reflects the sensitivity of the Shuvaev transform to the slope of the input PDFs through Eq.~\eqref{Shuvaevt}. The enhancement in the quark singlet GPD is clearly larger than that in the gluon one. However, our analysis shows that the contribution of the quarks is subleading and so the overall effect of incorporating the skewness through the Shuvaev transform is dictated by the gluon GPDs. Our results for the differences between $F^g(x,\xi,\mu_F)$ and $xg(x,\mu_F)$, and $F^{q,S}(x,\xi,\mu_F)$ and $q^S(x,\mu_F)$, are qualitatively similar to those presented in the DGLAP region at LO in Fig.~3 of Ref.~\cite{Bertone:2022frx}.

\section{Data-driven approach}
\label{sec:Datadriven}

As discussed in our previous works in the context of exclusive $J/\psi$ photoproduction in ${\rm Pb}+{\rm Pb}$ and O+O UPCs~\cite{Eskola:2022vaf,Eskola:2022vpi}, the photoproduction scattering amplitude $\mathcal M_A^{\gamma N \rightarrow \Upsilon N}(\xi,t=0)$ introduced above suffers from a large factorization/renormalization scale dependence. While it is milder for the case of $\Upsilon$ photoproduction considered here since the interaction scale is higher than in the $J/\psi$ photoproduction, it is still rather sizeable as we will show later on in Sec.~\ref{sec:Results}. In addition, the NLO results will be shown to somewhat underpredict the HERA and LHC  data on the $\gamma+p \to \Upsilon+p$ cross section. An approach to alleviate the strong scale dependence through consideration of additional power corrections $\sim \mathcal O(\mu_F^2/Q_0^2)$ arising in the so-called $Q_0$ subtraction, where $Q_0$ is the PDF or GPD parametrization scale, was advocated in~\cite{Jones:2016ldq,Flett:2019pux,Flett:2020duk,Flett:2021fvo,Flett:2022ues} in the context of $p+p$ and $p+{\rm Pb}$ collisions. Instead of the $Q_0$ subtraction,  we adopt a data-driven pQCD approach, where the $\gamma+{\rm Pb} \to \Upsilon+{\rm Pb}$
cross section is given by the product of the ratio between the $\Upsilon$ photoproduction cross sections on the nucleus and the proton calculated in NLO in pQCD, and the $\gamma+p \to \Upsilon+p$ cross section fitted to the available HERA~\cite{H1:2000kis,ZEUS:1998cdr,ZEUS:2009asc} and LHC data~\cite{LHCb:2015wlx},
\begin{equation}
\label{exp}
\sigma^{\gamma {\rm Pb} \rightarrow \Upsilon {\rm Pb}}(W) =  \left[ \frac { \sigma^{\gamma {\rm Pb} \rightarrow \Upsilon {\rm Pb}}(W)} { \sigma^{\gamma p \rightarrow \Upsilon p}(W)} \right]_{\rm pQCD}~\sigma_{{\rm fit}}^{\gamma p \rightarrow \Upsilon p}(W) \,.
\end{equation}
Using a simple power-like ansatz for $\sigma_{{\rm fit}}^{\gamma p \rightarrow \Upsilon p}(W)$ with an additional factor parametrizing the behavior of the 
cross section near the kinematic threshold~\cite{Guzey:2013xba}, one obtains~\cite{Kryshen:private}
\begin{align} 
\sigma_{{\rm fit}}^{\gamma p \rightarrow \Upsilon p}(W) & = \frac{0.902\ {\rm nb}\, {\rm GeV}^{-2}}{B_{\Upsilon}(W)} \left[1 - \frac{(M_{\Upsilon} + m_N)^2}{W^2} \right]^{1.5} \nonumber \\
& \times \left( \frac{W^2}{\widetilde{W}_0^2} \right)^{0.447} \,, 
\label{eq:fit}
\end{align}
with $\widetilde{W}_0 = 100$ GeV. Note that while the 2018 CMS data~\cite{CMS:2018bbk} have not been included in the fit, they are nevertheless well reproduced, see Fig.~\ref{fig:baseline} ahead. One way to interpret Eq.~(\ref{exp}) is that we supplement the fitted $\gamma+p \to \Upsilon+p$ cross sections by the theoretical nuclear modification $R(W)$,
\begin{equation}
\label{ratio}
R(W) = \left[ \frac { \sigma^{\gamma {\rm Pb} \rightarrow \Upsilon {\rm Pb}}(W) } { \sigma^{\gamma p \rightarrow \Upsilon p}(W)  } \right]_{\rm pQCD} \,,
\end{equation}
which can be anticipated to carry a reduced dependence on the choice of the factorization scale and on the explicit modeling of GPDs. In the first approximation, these effects cancel in $R(W)$ and it becomes mainly sensitive to the PDFs of protons and nuclei. Alternatively, one can interpret that in Eq.~(\ref{exp}) one rescales the calculated $\gamma+{\rm Pb} \to \Upsilon+{\rm Pb}$ cross sections by a factor that is needed to match the calculated $\gamma+p \to \Upsilon+p$ cross sections with the experimental ones -- an effective ``K factor''. In what follows, we will call the cross sections computed through Eq.~(\ref{exp}) the ``data-driven'' ones, in contrast to the ``standard'' pQCD predictions calculated without any reference to experimental data. The approach here is similar in spirit to the leading-order pQCD analysis of the nuclear suppression factor for exclusive $J/\psi$ photoproduction in ${\rm Pb}+{\rm Pb}$ collisions introduced and discussed in Refs.~\cite{Guzey:2013xba,Guzey:2013qza,Guzey:2020ntc}.  

\section{Results}
\label{sec:Results}

In this section, we present and discuss our results for the $\Upsilon$ photoproduction process on the proton, $\gamma+p \rightarrow \Upsilon+p$, and the rapidity-differential $\Upsilon$ spectra in ${\rm Pb}+{\rm Pb}$ UPCs, ${\rm Pb} + {\rm Pb} \rightarrow {\rm Pb} + \Upsilon + {\rm Pb}$. To estimate the sensitivity of our predictions to higher-order perturbative corrections, we adopt a standard, conservative prescription and vary the factorization and renormalization scales together in the interval of $\mu_{F} = \mu_{R} \in \left\{1/2, 1, 2 \right\} \times m_b$. As input proton and nuclear PDFs, we use CT18ANLO~\cite{Hou:2019efy} and EPPS21~\cite{Eskola:2021nhw} PDFs, respectively, from the LHAPDF interface~\cite{Buckley:2014ana}. The corresponding GPDs are obtained using the Shuvaev transform as discussed in Sec.~\ref{sec:Shuvaev}. Note that we use the version ``A'' of the CT18NLO analysis since this was the free proton baseline used in the EPPS21 nPDF analysis. It differs from the default CT18NLO mainly in the strange quark distributions. In the first instance we make NLO predictions following the standard pQCD approach, and then subsequently compare and contrast features of these predictions with those obtained from the data-driven method explained in Sec.~\ref{sec:Datadriven}, as well as with our earlier analyses \cite{Eskola:2022vpi,Eskola:2022vaf} of $J/\psi$ photoproduction in ${\rm Pb}+{\rm Pb}$ UPCs. 

\subsection{Standard pQCD results for $\gamma+p \to \Upsilon+p$ cross section}
\label{subsec:res_proton}

\begin{figure*}
\centering
\includegraphics[width=0.75\textwidth]{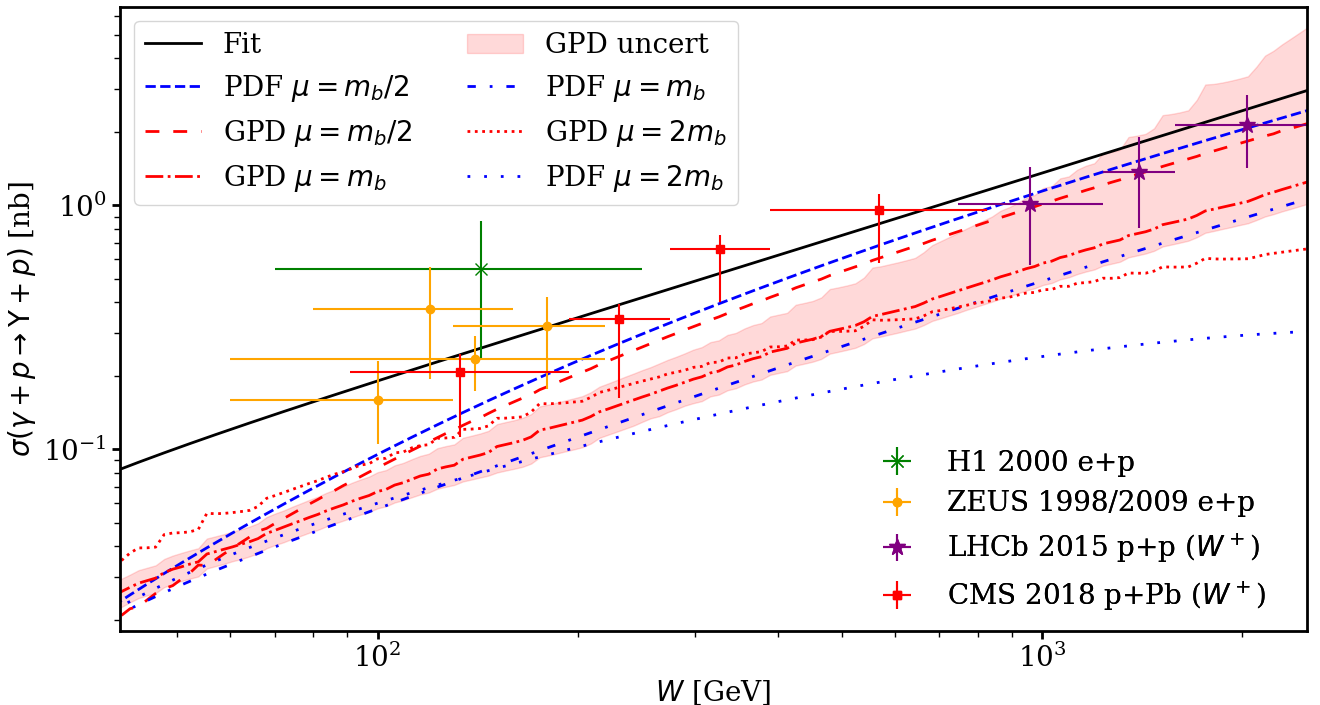}
\caption{The $\gamma+ p \rightarrow \Upsilon+ p$ cross section as a function of $W$. The NLO pQCD GPD-based (red curves) and PDF-based (blue curves) predictions evaluated at $\mu=\{m_b/2, m_b, 2m_b\}$ are presented by the dashed, dot-dashed and dotted lines; the shaded band is the propagated CT18ANLO PDF uncertainty for the GPD-based result at $\mu=m_b$. The HERA~\cite{ZEUS:1998cdr,H1:2000kis,ZEUS:2009asc} and LHC data~\cite{LHCb:2015wlx,CMS:2018bbk} data on this process are shown as well, together with a fit [Eq.~(\ref{eq:fit})] to the HERA data (the black solid line labeled ``Fit'').
}
\label{fig:baseline}
\end{figure*}

Figure~\ref{fig:baseline} presents the $\sigma^{\gamma p \to \Upsilon p}(W)$ cross section of exclusive $\Upsilon$ photoproduction on the proton, $\gamma+p \to \Upsilon+p$, as a function of the invariant photon-proton c.m.s. energy $W$. The dashed, dot-dashed and dotted curves correspond to the NLO pQCD predictions of Eq.~(\ref{eq:cs1_p}), which as input use either the proton GPDs obtained via the Shuvaev transform (the curves labeled ``GPD'') or the usual proton PDFs, i.e., the $\xi=0$ forward limit of the GPDs (the curves labeled ``PDF''). Each pair of predictions is evaluated with three scale settings $\mu = \mu_{F} = \mu_{R} \in \left\{ 1/2, 1, 2 \right\} \times m_b$. The shaded band represents the propagated uncertainty of the proton PDFs used for the GPD-based predictions at $\mu = m_b$. These results are compared with the available HERA~\cite{ZEUS:1998cdr,H1:2000kis,ZEUS:2009asc} and the LHC data~\cite{LHCb:2015wlx,CMS:2018bbk} on this process. Note that it is argued in~\cite{Flett:2021fvo} that the extracted values of $\sigma^{\gamma p \rightarrow \Upsilon p}(W)$ at the largest $W$ from the LHCb rapidity-differential measurements~\cite{LHCb:2015wlx} should be shifted upwards because the collaboration used a less accurate approximation for the photon flux in their analysis. Finally, the black solid line labeled ``Fit'' is the parametrization of Eq.~(\ref{eq:fit}).

One can see from the figure that while our NLO pQCD predictions reproduce the trends of the $W$ dependence of the data, they underestimate the normalization of the cross section, especially at larger values of $\mu$. A reliable description of the normalization would thus require a better theoretical understanding of the perturbative structure of the process including, e.g. the relevance of unknown next-to-NLO corrections, significance of the double logarithmic $\alpha_s \log (\mu_F^2/m_b^2) \log(1/\xi)$ terms present already in the NLO hard coefficient functions $T_g$ and $T_q$ \cite{Ivanov:2004vd,Ivanov:2015hca}\footnote{Note that these terms should be more relevant at low $\xi$ i.e. at high $W$ whereas the normalization seems to be an increasingly serious issue towards low values of $W$.}, and the size of the relativistic corrections to the quarkonium wave function~\cite{Lappi:2020ufv}. An account of these effects is beyond the scope of this paper and we will work around these issues through the data-driven predictions.

The systematics of the NLO pQCD predictions in Fig.~\ref{fig:baseline} can be summarized as follows. First, as discussed in Sec.~\ref{sec:Shuvaev}, the effect of skewness is rather mild, i.e., the difference between the GPD-based and PDF-based predictions is small, especially at smaller values of $\mu$. Second, while the GPD-based predictions correspond to higher values of $\sigma^{\gamma p \to \Upsilon p}(W)$ than the corresponding PDF-based ones at $\mu=m_b$ and $\mu=2m_b$, this hierarchy of predictions is reversed at $\mu=m_b/2$. A detailed examination indicates that this originates from a delicate interplay among the LO gluon and NLO gluon and quark contributions in $\mathcal M_A^{\gamma N \rightarrow \Upsilon N}(\xi, t=0)$ whose relative signs vary depending on the scale choices. This is further complicated by the fact that the magnitude of the skewness effect generated by the Shuvaev transform~(\ref{Shuvaevt}) depends on both $W$ (through its dependence on $\xi$) and $\mu_F$ controlling the slope of the $x$ dependence of the gluon and quark PDFs. Third, as a result of scale-dependent sign differences of quark/gluon contributions, the relative ordering of predictions from low to high $\mu$ depends on $W$.

\subsection{Standard pQCD results for ${\rm Pb} + {\rm Pb} \rightarrow {\rm Pb} + \Upsilon + {\rm Pb}$ UPC cross section}
\label{subsec:res_A}

In Fig.~\ref{spQCDdiff}, we show our standard NLO pQCD predictions for $\text{d}\sigma^{{\rm Pb} + {\rm Pb} \rightarrow {\rm Pb} + \Upsilon + {\rm Pb}}/\text{d}y$ as a function of the $\Upsilon$ rapidity $y$ at $\sqrt{s_{NN}} = 5.02$ TeV, see Eqs.~(\ref{XS_plus_minus}), (\ref{eq:cs1}), (\ref{eq:cs2}) and ({\ref{eq:M}). As input, we use the nuclear GPDs constructed using the Shuvaev transform and the EPPS21 nPDFs (central plus error sets). The three curves correspond to the three different choices of the factorization/renormalization scales $\mu=\{m_b/2, m_b, 2 m_b\}$. The shaded band gives the propagated uncertainty of the EPPS21 nPDFs in the $\mu=m_b$ case. As a useful reference, the upper $x$-axis shows the values of $W^+$ corresponding to each $y$, that is, $W^{+}=(M_{\Upsilon} \sqrt{s_{NN}} e^{y})^{1/2}$.

Two features of the results in Fig.~\ref{spQCDdiff} deserve to be mentioned. First, one can see from the figure that apart from the very tails of the rapidity distribution, $|y|>3$, the central prediction with $\mu = m_b$ does not lie between the other scale choice predictions with $\mu = m_b/2$ and $\mu = 2m_b$. This feature can be readily observed also in the results for the proton cross section in Fig.~\ref{fig:baseline}. Indeed, taking, for instance, $y=0$ corresponding to $W \approx 200$ GeV, one can see that the predictions for $\sigma^{\gamma p \rightarrow \Upsilon  p}(W)$ with $\mu=m_b$ lie below the corresponding predictions at $\mu_F=m_b/2$ and $\mu_F=2 m_b$.  Second, the scale uncertainty is rather large and the prediction with $\mu = m_b/2$ lies outside the nPDF uncertainty band. We will show in Sec.~\ref{subsec:data-driven} that both of these features can be tamed through our data-driven approach.

\begin{figure*}
    \centering
    \includegraphics[width=0.75\textwidth]{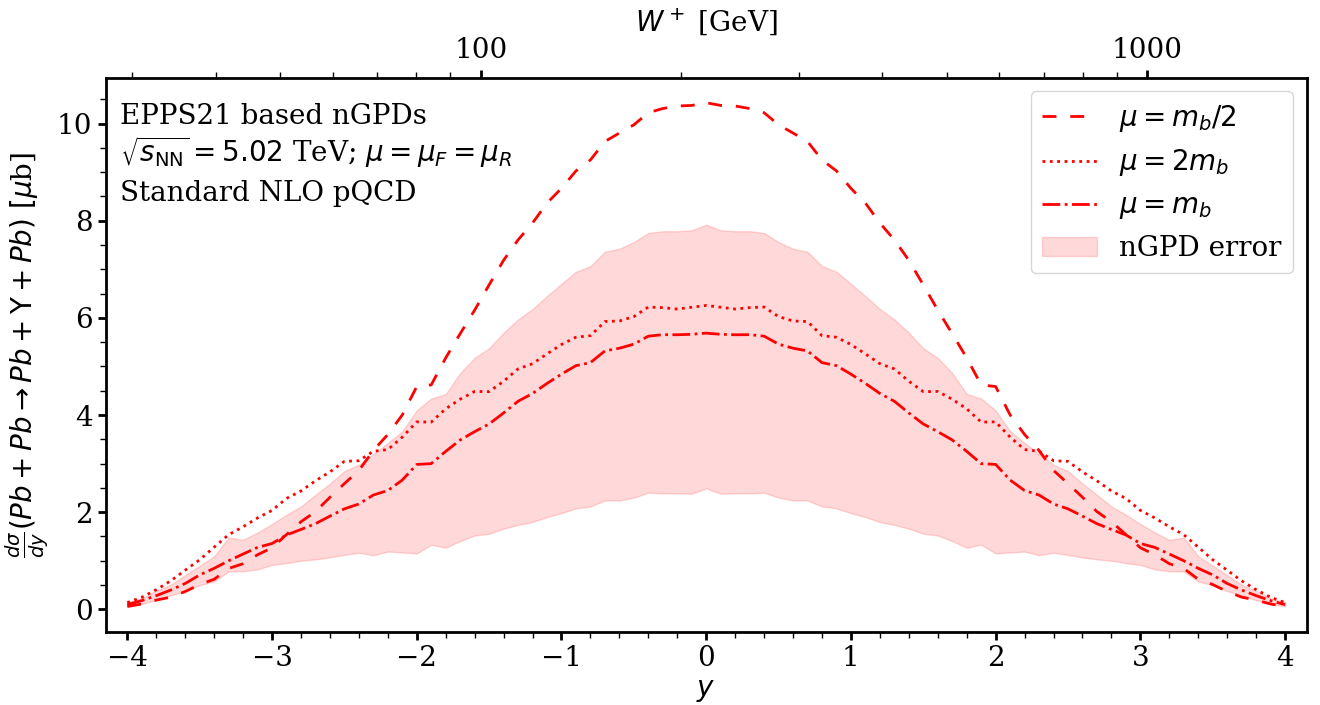}
    \caption{Standard NLO pQCD prediction for the rapidity-differential cross section for exclusive coherent $\Upsilon$ photoproduction in ${\rm Pb}+{\rm Pb}$ UPCs as a function of the $\Upsilon$ rapidity $y$ at $\sqrt{s_{NN}} = 5.02$ TeV. We use nuclear GPDs constructed from the EPPS21 nPDFs via the Shuvaev transform. The dashed-dotted curve represents the prediction with $\mu = m_b$, and the band indicates the nPDF-originating uncertainty evaluated at the same scale. The predictions with $\mu = m_b/2$ (dashed) and $\mu = 2m_b$ (dotted) are also shown. The upper $x$-axis shows the values of $W^+$ as a function of $y$.}
    \label{spQCDdiff}
\end{figure*}

In Figs.~\ref{qgi},~\ref{wcompts} and~\ref{reim}, we show various decompositions of $\text{d}\sigma^{{\rm Pb} + {\rm Pb} \rightarrow {\rm Pb} + \Upsilon + {\rm Pb}}/\text{d}y$ at $\mu = m_b$ as a function of $y$. Figure~\ref{qgi} presents the breakdown of the full cross section into the quark, gluon and interference contributions. It is clear that over the entire considered rapidity region, the gluon contribution dominates the quark contribution, in dissimilarity to the analogous breakdown for the $J/\psi$ rapidity-differential cross section in ${\rm Pb}+{\rm Pb}$ UPCs in NLO pQCD shown in our previous studies~\cite{Eskola:2022vaf,Eskola:2022vpi}, where the quark contribution was shown to be the dominant one around mid rapidity. One should note that 
even if the quark contribution is small, it is not zero or structureless and it leads to a visible contribution in the interference terms. One can speculate that the interaction scale in the $\Upsilon$ photoproduction is already sufficiently large so that NNLO corrections will not change the mutual hierarchy of quark/gluon contributions. The situation could be very different in the case of $J/\psi$ photoproduction where the quark dominance is a consequence of a coincidental cancellation between the LO and NLO gluon contributions.

\begin{figure*}
\centering
\includegraphics[width=0.7\textwidth]{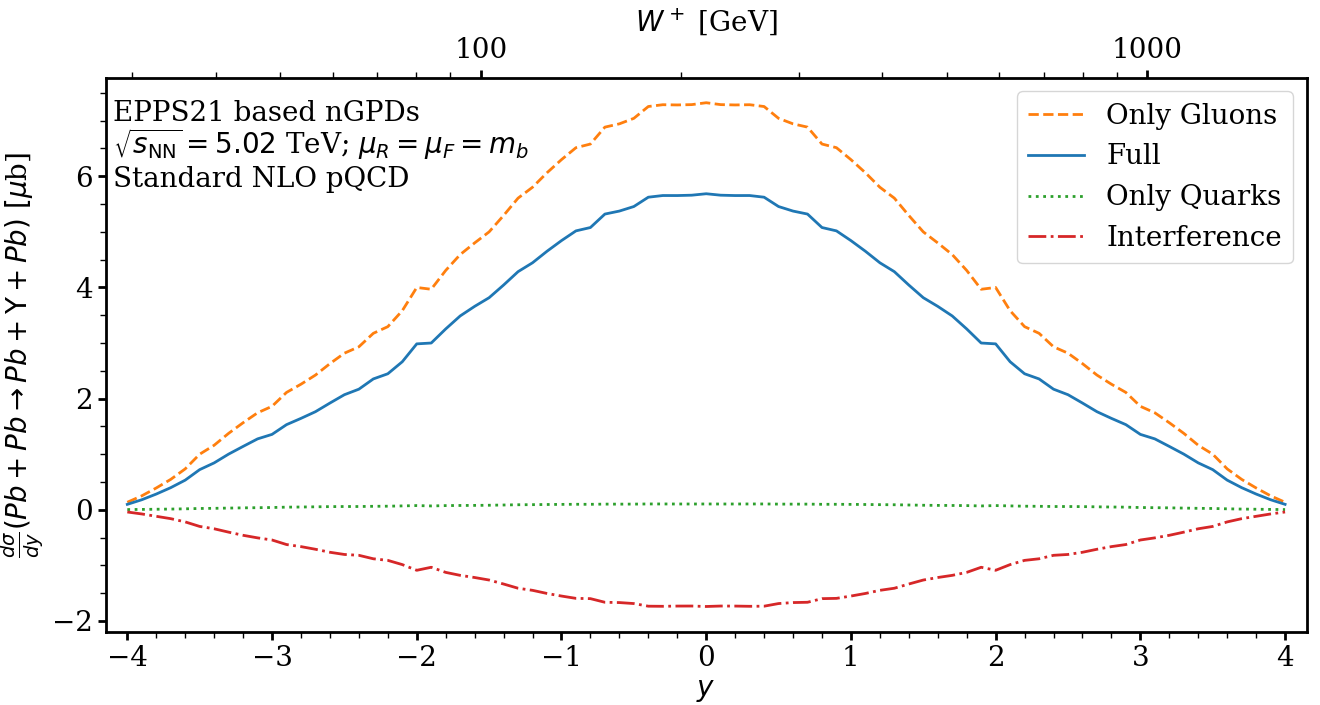}
   \caption{Decomposition of the rapidity-differential ${{\rm Pb} + {\rm Pb} \rightarrow {\rm Pb} + \Upsilon + {\rm Pb}}$ cross section with $\mu=m_b$ into the quark, gluon and quark-gluon interference contributions, in our standard NLO pQCD approach.}
    \label{qgi}
\end{figure*}

\begin{figure*}
    \centering
    \includegraphics[width=0.675\textwidth]{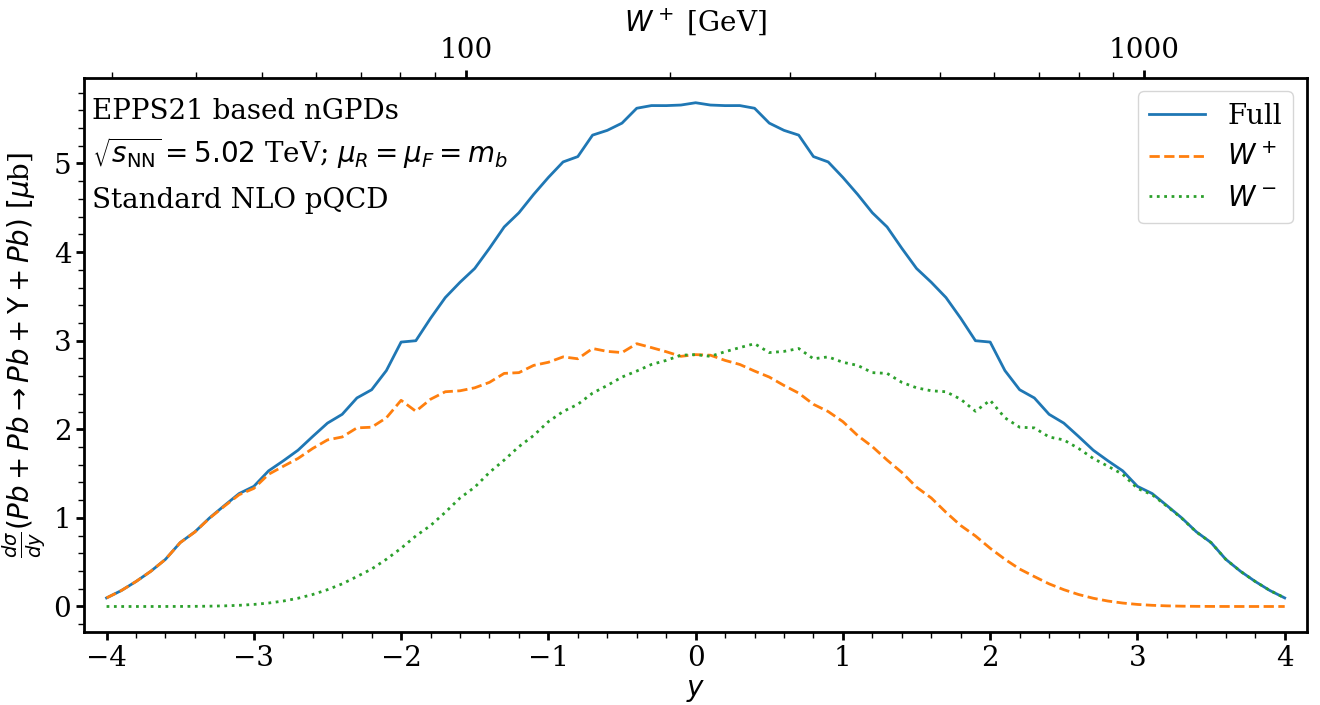}
    \caption{Decomposition of the rapidity-differential ${{\rm Pb} + {\rm Pb} \rightarrow {\rm Pb} + \Upsilon + {\rm Pb}}$ cross section with $\mu=m_b$ into the $W^+$ and $W^-$ components, in our standard NLO pQCD approach.}
    \label{wcompts}
\end{figure*}

In Fig.~\ref{wcompts}, we show the $W^+$ and $W^-$ decomposition i.e. separately plot the two contributions in Eq.~(\ref{XS_plus_minus}). The situation is very similar to that in our $J/\psi$ analysis, see~\cite{Eskola:2022vpi,Eskola:2022vaf} for more details. For instance, the $W^{-}$ contribution dominates at positive forward rapidities (large $W^{+}$) because there $(k dN_{\gamma}^{\rm Pb}/dk)_{k=k^{-}} \gg (k dN_{\gamma}^{\rm Pb}/dk)_{k=k^{+}}$. The situation is reversed in the region of backward rapidities corresponding to small $W^{+}$. The presence of two terms in Eq.~(\ref{XS_plus_minus}) complicates the extraction of the small-$x$ contribution from UPC cross sections at $y \neq 0$. However, it is possible to separate the $W^+$ and $W^-$ contributions by studying UPCs accompanied by forward neutron emission due to electromagnetic excitation of one or both colliding nuclei~\cite{Guzey:2013jaa}. Such an analysis in the case of coherent $J/\psi$ photoproduction in ${\rm Pb}+{\rm Pb}$ UPCs at 5.02 TeV was recently performed by the CMS collaboration~\cite{CMS:2022nnw}, which allowed one to deepen the small-$x$ reach down to $x \sim  10^{-4}$. 

Finally, Fig.~\ref{reim} presents the decomposition of \linebreak $\text{d}\sigma^{{\rm Pb} + {\rm Pb} \rightarrow {\rm Pb} + \Upsilon + {\rm Pb}}/\text{d}y$ into the contributions of the real and imaginary parts of $\mathcal M_A^{\gamma N \rightarrow \Upsilon N}(\xi, t=0)$. The imaginary part clearly dominates over the entire range of rapidity. Again, the situation was much more involved in the case of $J/\psi$, where the interplay of the two was highly non-trivial \cite{Eskola:2022vpi,Eskola:2022vaf}.

\begin{figure*}
    \centering
    \includegraphics[width=0.7\textwidth]{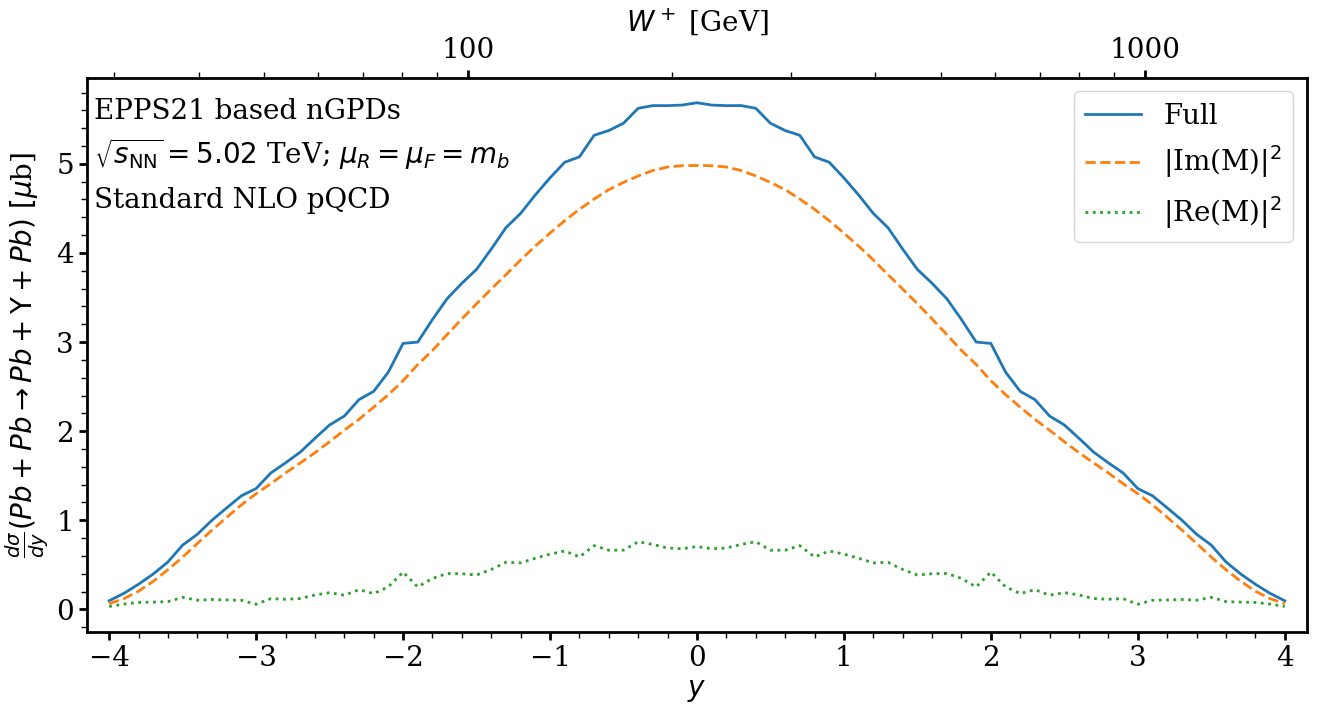}
    \caption{Decomposition of the rapidity-differential ${{\rm Pb} + {\rm Pb} \rightarrow {\rm Pb} + \Upsilon + {\rm Pb}}$ cross section with $\mu=m_b$ into the contributions from the real and imaginary parts, in our standard NLO pQCD approach.
    }
    \label{reim}
\end{figure*}

\subsection{Data-driven pQCD predictions for the ${\rm Pb} + {\rm Pb} \rightarrow {\rm Pb} + \Upsilon + {\rm Pb}$ UPC cross section}
\label{subsec:data-driven}

The data-driven pQCD prediction for the UPC cross section $\text{d}\sigma^{{\rm Pb} + {\rm Pb} \rightarrow {\rm Pb} + \Upsilon + {\rm Pb}}/\text{d}y$ is given by Eq.~(\ref{exp}), where only the ratio of the nucleus and proton cross sections $R(W)$ of Eq.~(\ref{ratio}) is calculated using our NLO pQCD framework, while the absolute normalization is given by $\sigma^{\gamma p \to \Upsilon p}_{\rm fit}(W)$ obtained from a fit to the proton data, see Eq.~(\ref{eq:fit}). The results for the differential cross section as a function of the $\Upsilon$ rapidity $y$ are shown in Fig.~\ref{fig:1}. The numerator and the denominator of the ratio $R(W)$ are calculated using the EPPS21-based and the CT18ANLO-based GPDs, respectively; these curves are labeled ``nGPD''. For comparison, we also show the results of the calculation, where we neglect the effect of skewness and use the forward $\xi \to 0$ limit for nuclear and proton GPDs; these curves are labeled ``nPDF''. The blue dot-dashed curve represents our central prediction at $\mu = m_b$ with the blue shaded band quantifying the propagation of the EPPS21 nPDF and the CT18ANLO proton PDF uncertainties; their counterparts in the case, where GPDs are taken in the forward limit, are given by the red solid curve and the corresponding red shaded band. The dotted and dashed curves correspond to the ratio $R(W)$ evaluated at $\mu = m_b/2$ and $\mu = 2m_b$, respectively. The uncertainties in $\sigma^{\gamma p \to \Upsilon p}_{\rm fit}(W)$ are not included in our estimates. For reference, we give the values of $W^{+} = (M_{\Upsilon} \sqrt{s_{NN}}\, e^{y})^{1/2}$ probed at a given rapidity $y$ on the upper $x$-axis, and also mark in the figure the points $|y| = 2$, beyond which the $\sigma^{\gamma p \to \Upsilon p}_{\rm fit}(W)$ fit to the $\gamma+ p \rightarrow \Upsilon + p$ photoproduction data is an extrapolation: the HERA data are available only for $W \geq 100 \,{\rm GeV}$ (see Fig.~\ref{fig:baseline}), but for $|y| \geq 2$ there is a large contribution from $W < 100 \,{\rm GeV}$, see Fig.~\ref{wcompts}.

It is important to contrast our results in Fig.~\ref{fig:1} with the standard NLO pQCD predictions shown in Fig.~\ref{spQCDdiff}. First, while the shapes of the $y$ distribution are very similar, the normalization of the data-driven results is approximately a factor of $2-2.5$ higher. This is a straightforward consequence of the rescaling of the cross section of exclusive $\Upsilon$ photoproduction on the proton to fit the available data. Second, the dependence on the factorization/renormalization scale $\mu$ is now more regular in the central rapidities: the central prediction with $\mu=m_b$ lies below the $\mu=m_b/2$ result and above the $\mu=2 m_b$ one. Most importantly, the scale dependence has reduced significantly. Third, the effects of GPD modeling are seen to largely cancel in the ratio $R(W)$. As a result, the data-driven pQCD predictions for $\text{d}\sigma^{{\rm Pb} + {\rm Pb} \rightarrow {\rm Pb} + \Upsilon + {\rm Pb}}/\text{d}y$ are here mainly sensitive to the input PDFs. Note that for lower $W$, where the real part restoration via Eq.~\eqref{re} is less accurate, the behavior of our results is less regular, but this lies in the tails of the $y$ distributions where our predictions anyhow lean on an extrapolation of $\sigma^{\gamma p \to \Upsilon p}_{\rm fit}(W)$ into non-measured values of $W$.

\begin{figure*}
    \centering
    \includegraphics[width=0.75\textwidth]{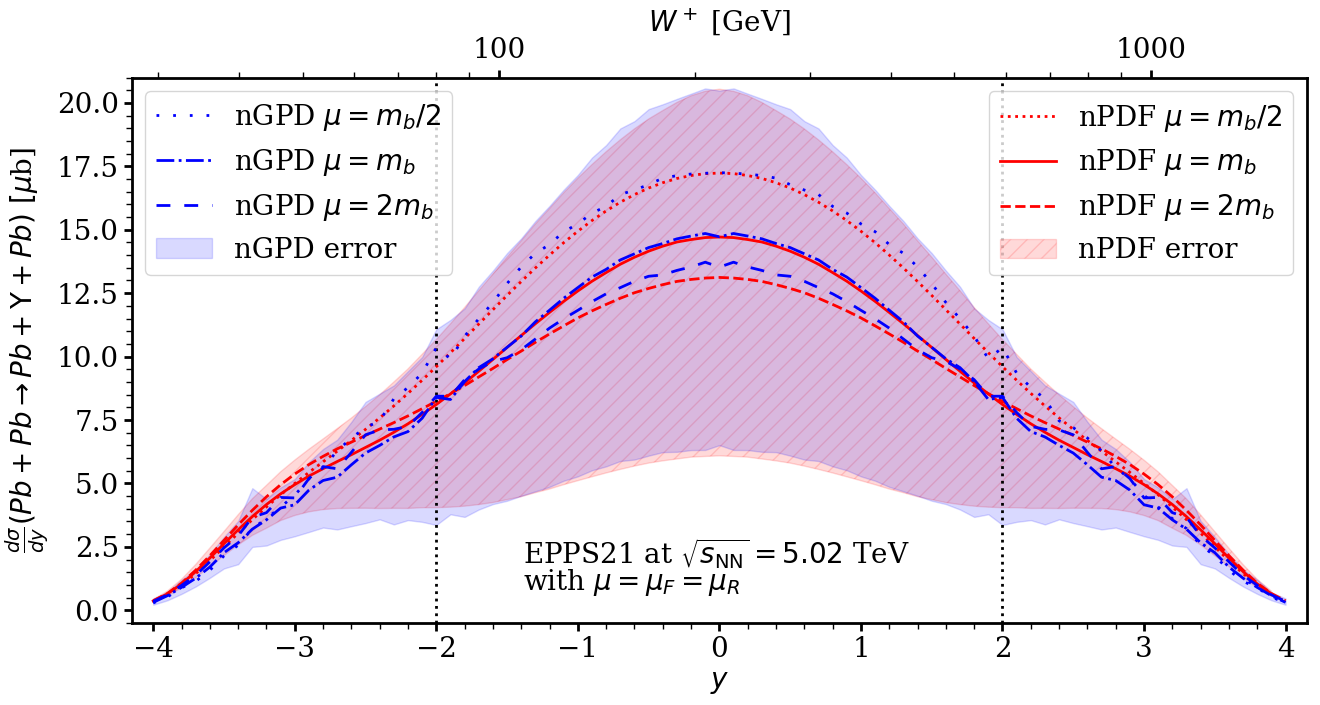}
    \caption{Data-driven NLO pQCD prediction for the rapidity-differential cross section for exclusive coherent $\Upsilon$ photoproduction in ${\rm Pb}+{\rm Pb}$ UPCs as a function of the $\Upsilon$ rapidity $y$ at $\sqrt{s_{NN}} = 5.02$ TeV. We use the nuclear and proton GPDs constructed from the EPPS21 nPDFs and CT18ANLO proton PDFs, respectively, obtained via the Shuvaev transform (the curves labeled ``nGPD'').  For comparison, we also show the results based on the $\xi=0$ limit of the used GPDs (the curves labeled ``nPDF''). The blue dot-dashed line represents the central prediction with $\mu = m_b$ and the blue band gives the propagated uncertainties of the nuclear and proton PDFs. The predictions for $\mu = m_b/2$ (dotted) and $\mu = 2m_b$ (dashed) are also shown. The upper $x$-axis shows the values of $W^+$ for each $y$. The vertical dashed lines denote the points $|y| = 2$, beyond which the results are sensitive to low $W$ where $\sigma^{\gamma p \to \Upsilon p}_{\rm fit}(W)$ is an extrapolation.
}
\label{fig:1}
\end{figure*}

To quantify the magnitude of nuclear effects probed in exclusive $\Upsilon$ photoproduction in ${\rm Pb}+{\rm Pb}$ UPCs at the LHC, it is convenient to consider separately the ratio $R(W)$ in Eq.~(\ref{ratio}). Indeed, at a given value of the $\Upsilon$ rapidity $y \neq 0$, the ${\rm Pb}+{\rm Pb}$ UPC cross section contains two terms leading to a two-fold ambiguity in the photon-nucleon c.m.s.~energy $W^{\pm}$. As a consequence, this mixes the low-$x$ and medium-$x$ contributions to the UPC cross section and makes it challenging to extract the information on small-$x$ physics, which is often thought to be at the heart of the process under consideration. This issue is absent in the case of $R(W)$ although it cannot be experimentally measured in a model-independent way. In the upper panel of Fig.~\ref{fig:r}, we show the ratio $R(W)$ as a function of $W^+$. On the $x$-axis at the top, we also give the corresponding values of the skewness $\xi^{+} = M_{\Upsilon}^2/[2(W^{+})^2-M_{\Upsilon}^2]$. The curve corresponds to the central prediction at $\mu = m_b$ shown in Fig.~\ref{fig:1}, where the numerator and the denominator of $R(W)$ are calculated using the EPP21-based nuclear GPDs and the CT18ANLO-based free proton GPDs, respectively. The shaded band is the result of the propagation of the EPPS21 nPDF and the CT18ANLO proton PDF uncertainties. We see that the rescaling factor $R(W)$ depends strongly on $W$ and its value can be as large as several hundreds. The absolute value can, however, be mostly explained through the proton and nuclear form factors. To see this and to provide a closer comparison with nuclear modifications of nPDFs, one can eliminate the effects of the nuclear and the proton form factors in the $R(W)$ ratio by rescaling it by the factor of $R^{\prime}(W)$,
\begin{equation}
R^{\prime}(W)= \frac{1/B_{\Upsilon}(W)}{\int_{|t_{\rm min}|}^{\infty} \text{d}t \, |F_A(-t)|^2} \,,
\label{eq:R_prime}
\end{equation}
where $B_{\Upsilon}(W)$ is the slope of the $t$ dependence of the $\gamma+p \to \Upsilon+p$ differential cross section in Eq.~(\ref{eq:B_slope}) and $F_A(t)$ is the nuclear form factor in Eq.~(\ref{eq:F_A}). Note that $R^{\prime}(W)$ depends on $W^{+}$ through $|t_{\rm min}|=m_N^2 (M_{\Upsilon}/W^{+})^4$ and $B_{\Upsilon}(W^+)$. In the lower panel of Fig.~\ref{fig:r}, we present the scaled $R(W)$ ratio, i.e., the product $R(W) \times R^{\prime}(W)$, as a function of $W^{+}$ by the red solid curve. The propagated nuclear and free proton PDF uncertainties are given by the red shaded band. One can see from the figure that as a function of $\xi^+$,  $R(W) \times R^{\prime}(W)$ exhibits significant suppression for small $\xi^+ < 0.05$ and a $\sim 10$\% enhancement at $\xi^+ \sim 0.1$. This behaviour reflects the characteristic nuclear modifications of nPDFs associated with nuclear shadowing at small $x$ and nuclear anti-shadowing at $x \sim 0.1$. To highlight the latter point, we also show the squared EPPS21 nuclear modification factors for the gluon and quark singlet, 
\begin{align}
R_g^2(\xi,\mu_F) & =\left[\frac{g_A(\xi,\mu_F)}{g_p(\xi,\mu_F)}\right]^2 \,, \label{eq:RG2} \\
R_q^2(\xi,\mu_F) & =\left[\frac{q^{S}_A(\xi,\mu_F)}{q^{S}_p(\xi,\mu_F)}\right]^2 \,, \label{eq:RQ2}
\end{align}
as a function of $\xi=\xi^{+}$, where $g_A$ ($q_A^{S}$) and $g_p$ ($q^S_p$) are the gluon (quark-singlet) distributions per nucleon in the nucleus and the free proton, respectively. The corresponding shaded bands represent the EPPS21 nPDF uncertainties of these ratios. One can see that the shape and normalization of both $R_g^2(\xi)$ and $R_q^2(\xi)$ is similar to those of $R(W) \times R^{\prime}(W)$.
Moreover, because of the dominance of the gluon-initiated contribution over the quark one, see Fig.~\ref{qgi}, and the flat shape of the gluon nuclear modifications at small $x$, the values of $R(W) \times R^{\prime}(W)$ and $R_g^2(\xi)$ become very close for $\xi^{+}~\leq~10^{-3}$ ($W^{+}~>~200$~GeV).

\begin{figure*}
    \centering
    \includegraphics[width=0.75\textwidth]{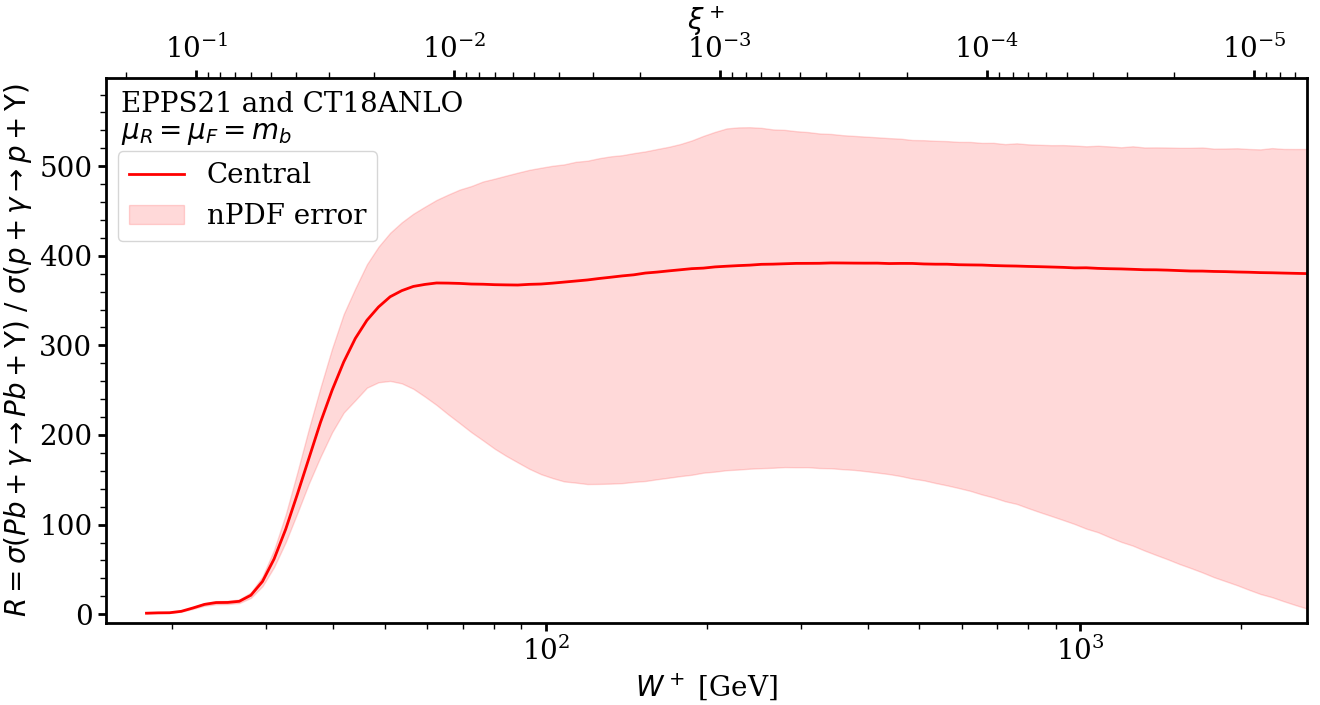}
    \includegraphics[width=0.75\textwidth]{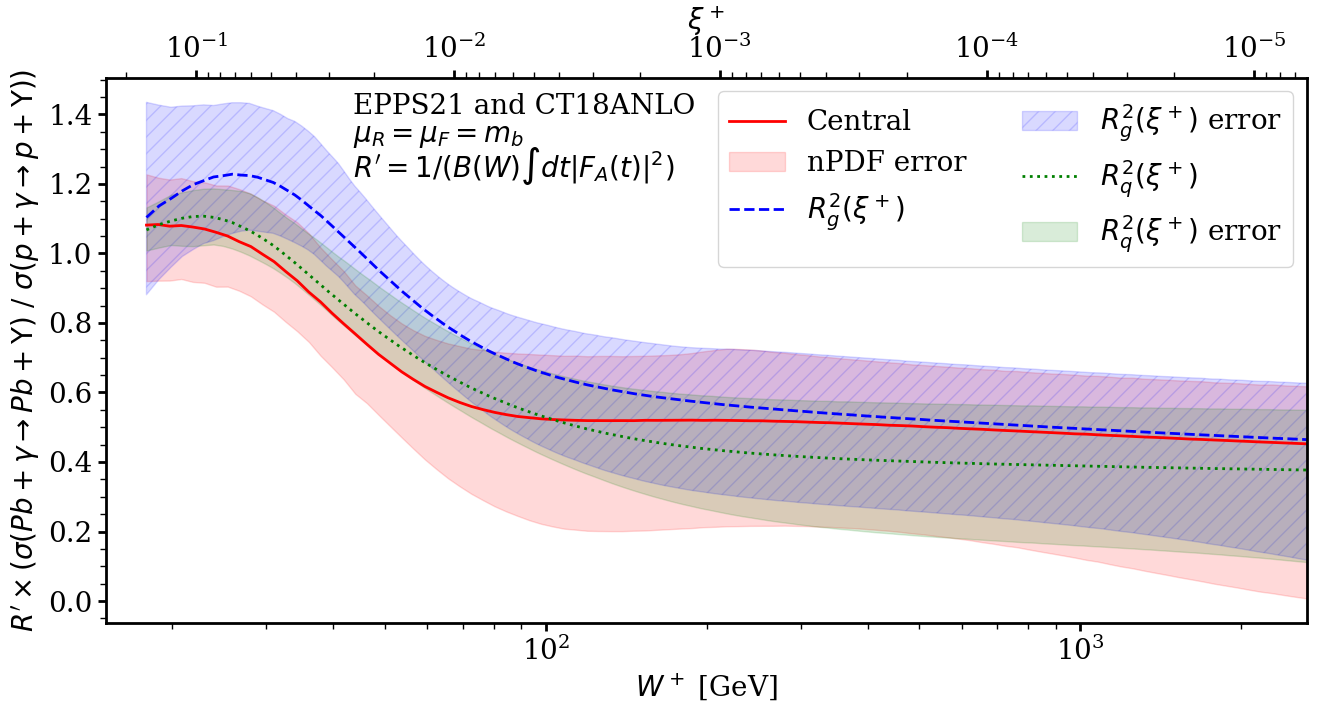}
    \caption{
        Upper panel: The ratio $R(W) = \left[\sigma^{\gamma {\rm Pb} \rightarrow \Upsilon {\rm Pb}}(W)/ \sigma^{\gamma p \rightarrow \Upsilon p}(W)\right]_{\rm pQCD}$ as a function of the c.m.s. energy $W^+$ evaluated using the EPPS21 nuclear and CT18ANLO free proton PDFs at $\mu = m_b$. The shaded band corresponds to the EPPS21 and CT18ANLO PDFs uncertainties. The upper 
        $x$-axis indicates the corresponding values of the skewness $\xi^+$. Lower panel: the rescaled ratio $R^{\prime}(W) \times R(W)$ as a function of $W^{+}$. For comparison, the EPPS21 gluon and quark-singlet nuclear modifications squared along with their uncertainties are overlaid. The shaded bands show the PDF-originated uncertainties.    
    }
    \label{fig:r}
\end{figure*}

\subsection{Feasibility of the measurement of $\Upsilon$ photoproduction in ${\rm Pb}+{\rm Pb}$ UPCs at the LHC}
\label{subsec:feas}

Having now obtained an educated estimate for the $\Upsilon$ cross section in ${\rm Pb}+{\rm Pb}$ collisions, we will here check to what extent an experimental measurement of the process would be feasible. To this end, we lean on the exclusive $\Upsilon$ $p+{\rm Pb}$ measurement by the CMS collaboration \cite{CMS:2018bbk} at $\sqrt{s_{NN}} = 5.02\,{\rm TeV}$. This measurement with an integrated luminosity of $\mathcal L(p+\text{Pb}) =$ 32.6 nb$^{-1}$ reported $\sim 80$ identified $\Upsilon(1S)$ particles and yielded a total cross section $\sigma(p+\text{Pb}) = 94.8$~nb in the rapidity interval $|y| < 2.2$. If we desire a ${\rm Pb}+{\rm Pb}$ measurement that is as precise as the $p+{\rm Pb}$ measurement (the same number of events), and assume the same efficiency, the condition is 
\begin{equation}
  \sigma(p+\text{Pb})\mathcal L(p+\text{Pb})  = \sigma(\text{Pb}+\text{Pb}) \mathcal L(\text{Pb}+\text{Pb}) \,.
\end{equation}
From Fig.~\ref{fig:1}, we find a total cross section $\sigma(\text{Pb}+\text{Pb}) \sim 52\,\mu$b in the same rapidity interval $-2.2 < y < 2.2$.  It then follows that the required integrated luminosity should be
\begin{equation}
 \mathcal L(\text{Pb}+\text{Pb}) = 0.06\,\text{nb}^{-1} \,,
\end{equation}
to observe $\sim80$ events. Given that the recorded luminosity at the 2018 ${\rm Pb}+{\rm Pb}$ run for CMS is as high as 1.7 nb$^{-1}$~\cite{CMS:lumi}, our counting would thus promise $\sim 80 \times (1.7/0.06) \approx 2300$ events. Moreover, in Run III CMS aims for an integrated luminosity of 13 nb$^{-1}$ \cite{Citron:2018lsq}, so the measurement of exclusive $\Upsilon$ photoproduction in ${\rm Pb}+{\rm Pb}$ collisions looks more than feasible to be performed at the CMS experiment.

\section{Conclusions and Outlook}
\label{sec:Conclusions}

We presented the first study of the rapidity-differential cross section of exclusive $\Upsilon$ photoproduction in ultraperipheral lead-lead collisions at the LHC using collinear factorization at NLO pQCD.  In addition, we extended our previous framework in~\cite{Eskola:2022vaf,Eskola:2022vpi} by now including explicit GPD modeling through the Shuvaev integral transform. In our standard NLO pQCD approach, we showed that the GPD effects are small, and unlike in the $J/\psi$ case, the imaginary part and gluon contributions dominate the amplitude. The scale uncertainties are significantly reduced from the $J/\psi$ case, but they are still alarmingly large. In the $\gamma+p$ case, the NLO calculation was shown to underpredict the HERA data, which calls for further improvements such as NNLO pQCD and NRQCD corrections.

Using the $\gamma+ p \rightarrow \Upsilon + p$ cross section from HERA and the LHC as a baseline, we proposed a data-driven pQCD approach to make more constrained predictions for $\text{d}\sigma ({\rm Pb} + {\rm Pb} \rightarrow {\rm Pb} + \Upsilon + {\rm Pb})/\text{d}y$ and showed that the resulting factorization/renormalization dependence becomes smaller than that in the standard pQCD result for this process. In addition, effects due to the explicit GPD modeling largely cancel and most of the remaining uncertainty is due to PDFs of free and bound nucleons. This serves as a first step towards being able to include heavy quarkonia UPC data in the global analyses of nPDFs to provide constraints on partons inside nuclei at moderate to low $x$. We also estimated that the production cross sections are high enough for this process to be measured in ${\rm Pb}+{\rm Pb}$ collisions at the LHC. While the theoretical situation nevertheless seems a little better for the $\Upsilon$ production, the experimental statistics obtainable may be sparser than that for $J/\psi$. Future works can therefore include applying our data-driven approach to exclusive $J/\psi$ photoproduction in nucleus-nucleus collisions, where also the statistical quality of the baseline $\gamma + p$ data is greater than for $\Upsilon$ production. In the $J/\psi$ case, the GPD modeling given by the Shuvaev transform is surmised to have an even smaller effect in comparison to the $\Upsilon$ photoproduction considered here, but to what extent the scale dependence can be tamed, calls for a detailed analysis.

\vspace{-0.5cm}

\section*{Acknowledgments}

We acknowledge the financial support from the Magnus Ehrnrooth foundation (T.L.), the Academy of Finland
Projects No. 308301 (H.P.) and No. 330448 (K.J.E.). This research was funded as a part of the Center of Excellence in Quark Matter of the Academy of Finland (Projects No. 346325 and No. 346326). This research is part of the European Research Council Project No.
ERC-2018-ADG-835105 YoctoLHC.


\vspace{-0.5cm}

\begin{thebibliography}{99}

\bibitem{Bertulani:2005ru}
C.~A.~Bertulani, S.~R.~Klein and J.~Nystrand,
Ann. Rev. Nucl. Part. Sci. \textbf{55} (2005), 271-310
doi:10.1146/annurev.nucl.55.090704.151526
[arXiv:nucl-ex/0502005 [nucl-ex]].

\bibitem{Baltz:2007kq}
A.~J.~Baltz, G.~Baur, D.~d'Enterria, L.~Frankfurt, F.~Gelis, V.~Guzey, K.~Hencken, Y.~Kharlov, M.~Klasen and S.~R.~Klein, \textit{et al.}
Phys. Rept. \textbf{458} (2008), 1-171
doi:10.1016/j.physrep.2007.12.001
[arXiv:0706.3356 [nucl-ex]].

\bibitem{Contreras:2015dqa}
J.~G.~Contreras and J.~D.~Tapia Takaki,
Int. J. Mod. Phys. A \textbf{30} (2015), 1542012
doi:10.1142/S0217751X15420129

\bibitem{Klein:2019qfb}
S.~R.~Klein and H.~M\"antysaari,
Nature Rev. Phys. \textbf{1} (2019) no.11, 662-674
doi:10.1038/s42254-019-0107-6
[arXiv:1910.10858 [hep-ex]].

\bibitem{Ryskin:1992ui}
M.~G.~Ryskin,
Z. Phys. C \textbf{57} (1993), 89-92
doi:10.1007/BF01555742

\bibitem{Ivanov:2004vd}
D.~Y.~Ivanov, A.~Schafer, L.~Szymanowski and G.~Krasnikov,
Eur. Phys. J. C \textbf{34} (2004) no.3, 297-316
[erratum: Eur. Phys. J. C \textbf{75} (2015) no.2, 75]
doi:10.1140/epjc/s2004-01712-x
[arXiv:hep-ph/0401131 [hep-ph]].

\bibitem{Collins:1996fb}
J.~C.~Collins, L.~Frankfurt and M.~Strikman,
Phys. Rev. D \textbf{56} (1997), 2982-3006
doi:10.1103/PhysRevD.56.2982
[arXiv:hep-ph/9611433 [hep-ph]].

\bibitem{Ji:1996nm}
X.~D.~Ji,
Phys. Rev. D \textbf{55} (1997), 7114-7125
doi:10.1103/PhysRevD.55.7114
[arXiv:hep-ph/9609381 [hep-ph]].

\bibitem{Radyushkin:1997ki}
A.~V.~Radyushkin,
Phys. Rev. D \textbf{56} (1997), 5524-5557
doi:10.1103/PhysRevD.56.5524
[arXiv:hep-ph/9704207 [hep-ph]].

\bibitem{Diehl:2003ny}
M.~Diehl,
Phys. Rept. \textbf{388} (2003), 41-277
doi:10.1016/j.physrep.2003.08.002
[arXiv:hep-ph/0307382 [hep-ph]].

\bibitem{PHENIX:2009xtn}
S.~Afanasiev \textit{et al.} [PHENIX],
Phys. Lett. B \textbf{679} (2009), 321-329
doi:10.1016/j.physletb.2009.07.061
[arXiv:0903.2041 [nucl-ex]].

\bibitem{ALICE:2012yye}
B.~Abelev \textit{et al.} [ALICE],
Phys. Lett. B \textbf{718} (2013), 1273-1283
doi:10.1016/j.physletb.2012.11.059
[arXiv:1209.3715 [nucl-ex]].

\bibitem{ALICE:2013wjo}
E.~Abbas \textit{et al.} [ALICE],
Eur. Phys. J. C \textbf{73} (2013) no.11, 2617
doi:10.1140/epjc/s10052-013-2617-1
[arXiv:1305.1467 [nucl-ex]].

\bibitem{CMS:2016itn}
V.~Khachatryan \textit{et al.} [CMS],
Phys. Lett. B \textbf{772} (2017), 489-511
doi:10.1016/j.physletb.2017.07.001
[arXiv:1605.06966 [nucl-ex]].

\bibitem{ALICE:2019tqa}
S.~Acharya \textit{et al.} [ALICE],
Phys. Lett. B \textbf{798} (2019), 134926
doi:10.1016/j.physletb.2019.134926
[arXiv:1904.06272 [nucl-ex]].

\bibitem{ALICE:2021gpt}
S.~Acharya \textit{et al.} [ALICE],
Eur. Phys. J. C \textbf{81} (2021) no.8, 712
doi:10.1140/epjc/s10052-021-09437-6
[arXiv:2101.04577 [nucl-ex]].

\bibitem{LHCb:2021bfl}
R.~Aaij \textit{et al.} [LHCb],
JHEP \textbf{07} (2022), 117
doi:10.1007/JHEP07(2022)117
[arXiv:2107.03223 [hep-ex]].

\bibitem{LHCb:2022ahs}
LHCb Collaboration,
[arXiv:2206.08221 [hep-ex]].

\bibitem{Citron:2018lsq}
Z.~Citron, A.~Dainese, J.~F.~Grosse-Oetringhaus, J.~M.~Jowett, Y.~J.~Lee, U.~A.~Wiedemann, M.~Winn, A.~Andronic, F.~Bellini and E.~Bruna, \textit{et al.}
CERN Yellow Rep. Monogr. \textbf{7} (2019), 1159-1410
doi:10.23731/CYRM-2019-007.1159
[arXiv:1812.06772 [hep-ph]].

\bibitem{Kusina:2020lyz}
A.~Kusina, T.~Je\v{z}o, D.~B.~Clark, P.~Duwent\"aster, E.~Godat, T.~J.~Hobbs, J.~Kent, M.~Klasen, K.~Kova\v{r}\'\i{}k and F.~Lyonnet, \textit{et al.}
Eur. Phys. J. C \textbf{80} (2020) no.10, 968
doi:10.1140/epjc/s10052-020-08532-4
[arXiv:2007.09100 [hep-ph]].

\bibitem{Eskola:2021nhw}
K.~J.~Eskola, P.~Paakkinen, H.~Paukkunen and C.~A.~Salgado,
Eur. Phys. J. C \textbf{82} (2022) no.5, 413
doi:10.1140/epjc/s10052-022-10359-0
[arXiv:2112.12462 [hep-ph]].

\bibitem{Helenius:2021tof}
I.~Helenius, M.~Walt and W.~Vogelsang,
Phys. Rev. D \textbf{105} (2022) no.9, 9
doi:10.1103/PhysRevD.105.094031
[arXiv:2112.11904 [hep-ph]].

\bibitem{AbdulKhalek:2022fyi}
R.~Abdul Khalek, R.~Gauld, T.~Giani, E.~R.~Nocera, T.~R.~Rabemananjara and J.~Rojo,
Eur. Phys. J. C \textbf{82} (2022) no.6, 507
doi:10.1140/epjc/s10052-022-10417-7
[arXiv:2201.12363 [hep-ph]].

\bibitem{Eskola:2022vpi}
K.~J.~Eskola, C.~A.~Flett, V.~Guzey, T.~L\"oyt\"ainen and H.~Paukkunen,
Phys. Rev. C \textbf{106} (2022) no.3, 035202
doi:10.1103/PhysRevC.106.035202
[arXiv:2203.11613 [hep-ph]].

\bibitem{Eskola:2022vaf}
K.~J.~Eskola, C.~A.~Flett, V.~Guzey, T.~L\"oyt\"ainen and H.~Paukkunen,
[arXiv:2210.16048 [hep-ph]].

\bibitem{H1:2000kis}
C.~Adloff \textit{et al.} [H1],
Phys. Lett. B \textbf{483} (2000), 23-35
doi:10.1016/S0370-2693(00)00530-X
[arXiv:hep-ex/0003020 [hep-ex]].

\bibitem{ZEUS:1998cdr}
J.~Breitweg \textit{et al.} [ZEUS],
Phys. Lett. B \textbf{437} (1998), 432-444
doi:10.1016/S0370-2693(98)01081-8
[arXiv:hep-ex/9807020 [hep-ex]].

\bibitem{ZEUS:2009asc}
S.~Chekanov \textit{et al.} [ZEUS],
Phys. Lett. B \textbf{680} (2009), 4-12
doi:10.1016/j.physletb.2009.07.066
[arXiv:0903.4205 [hep-ex]].

\bibitem{LHCb:2015wlx}
R.~Aaij \textit{et al.} [LHCb],
JHEP \textbf{09} (2015), 084
doi:10.1007/JHEP09(2015)084
[arXiv:1505.08139 [hep-ex]].

\bibitem{CMS:2018bbk}
A.~M.~Sirunyan \textit{et al.} [CMS],
Eur. Phys. J. C \textbf{79} (2019) no.3, 277
[erratum: Eur. Phys. J. C \textbf{82} (2022) no.4, 343]
doi:10.1140/epjc/s10052-019-6774-8
[arXiv:1809.11080 [hep-ex]].

\bibitem{Shuvaev:1999fm}
A.~Shuvaev,
Phys. Rev. D \textbf{60} (1999), 116005
doi:10.1103/PhysRevD.60.116005
[arXiv:hep-ph/9902318 [hep-ph]].

\bibitem{Shuvaev:1999ce}
A.~G.~Shuvaev, K.~J.~Golec-Biernat, A.~D.~Martin and M.~G.~Ryskin,
Phys. Rev. D \textbf{60} (1999), 014015
doi:10.1103/PhysRevD.60.014015
[arXiv:hep-ph/9902410 [hep-ph]].

\bibitem{Golec-Biernat:1999trj}
K.~J.~Golec-Biernat, A.~D.~Martin and M.~G.~Ryskin,
Phys. Lett. B \textbf{456} (1999), 232-239
doi:10.1016/S0370-2693(99)00504-3
[arXiv:hep-ph/9903327 [hep-ph]].

\bibitem{Vidovic:1992ik}
M.~Vidovic, M.~Greiner, C.~Best and G.~Soff,
Phys. Rev. C \textbf{47} (1993), 2308-2319
doi:10.1103/PhysRevC.47.2308

\bibitem{Woods:1954zz}
R.~D.~Woods and D.~S.~Saxon,
Phys. Rev. \textbf{95} (1954), 577-578
doi:10.1103/PhysRev.95.577

\bibitem{Hoodbhoy:1996zg}
P.~Hoodbhoy,
Phys. Rev. D \textbf{56} (1997), 388-393
doi:10.1103/PhysRevD.56.388
[arXiv:hep-ph/9611207 [hep-ph]].

\bibitem{Ryskin:1995hz}
M.~G.~Ryskin, R.~G.~Roberts, A.~D.~Martin and E.~M.~Levin,
Z. Phys. C \textbf{76} (1997), 231-239
doi:10.1007/s002880050547
[arXiv:hep-ph/9511228 [hep-ph]].

\bibitem{H1:2013okq}
C.~Alexa \textit{et al.} [H1],
Eur. Phys. J. C \textbf{73} (2013) no.6, 2466
doi:10.1140/epjc/s10052-013-2466-y
[arXiv:1304.5162 [hep-ex]].

\bibitem{Khoze:2013dha}
V.~A.~Khoze, A.~D.~Martin and M.~G.~Ryskin,
Eur. Phys. J. C \textbf{73} (2013), 2503
doi:10.1140/epjc/s10052-013-2503-x
[arXiv:1306.2149 [hep-ph]].

\bibitem{Berthou:2015oaw}
B.~Berthou, D.~Binosi, N.~Chouika, L.~Colaneri, M.~Guidal, C.~Mezrag, H.~Moutarde, J.~Rodr\'\i{}guez-Quintero, F.~Sabati\'e and P.~Sznajder, \textit{et al.}
Eur. Phys. J. C \textbf{78} (2018) no.6, 478
doi:10.1140/epjc/s10052-018-5948-0
[arXiv:1512.06174 [hep-ph]].

\bibitem{Martin:2008gqx}
A.~D.~Martin, C.~Nockles, M.~G.~Ryskin, A.~G.~Shuvaev and T.~Teubner,
Eur. Phys. J. C \textbf{63} (2009), 57-67
doi:10.1140/epjc/s10052-009-1087-y
[arXiv:0812.3558 [hep-ph]].

\bibitem{Kumericki:2009uq}
K.~Kumeri\v{c}ki and D.~Mueller,
Nucl. Phys. B \textbf{841} (2010), 1-58
doi:10.1016/j.nuclphysb.2010.07.015
[arXiv:0904.0458 [hep-ph]].

\bibitem{Bertone:2022frx}
V.~Bertone, H.~Dutrieux, C.~Mezrag, J.~M.~Morgado and H.~Moutarde,
Eur. Phys. J. C \textbf{82} (2022) no.10, 888
doi:10.1140/epjc/s10052-022-10793-0
[arXiv:2206.01412 [hep-ph]].

\bibitem{Dutrieux:2023qnz}
H.~Dutrieux, M.~Winn and V.~Bertone,
[arXiv:2302.07861 [hep-ph]].

\bibitem{Hou:2019efy}
T.~J.~Hou, J.~Gao, T.~J.~Hobbs, K.~Xie, S.~Dulat, M.~Guzzi, J.~Huston, P.~Nadolsky, J.~Pumplin and C.~Schmidt, \textit{et al.}
Phys. Rev. D \textbf{103} (2021) no.1, 014013
doi:10.1103/PhysRevD.103.014013
[arXiv:1912.10053 [hep-ph]].

\bibitem{Buckley:2014ana}
A.~Buckley, J.~Ferrando, S.~Lloyd, K.~Nordstr\"om, B.~Page, M.~R\"ufenacht, M.~Sch\"onherr and G.~Watt,
Eur. Phys. J. C \textbf{75} (2015), 132
doi:10.1140/epjc/s10052-015-3318-8
[arXiv:1412.7420 [hep-ph]].

\bibitem{Diehl:2007zu}
M.~Diehl and W.~Kugler,
Phys. Lett. B \textbf{660} (2008), 202-211
doi:10.1016/j.physletb.2007.12.047
[arXiv:0711.2184 [hep-ph]].

\bibitem{Jones:2016ldq}
S.~P.~Jones, A.~D.~Martin, M.~G.~Ryskin and T.~Teubner,
Eur. Phys. J. C \textbf{76} (2016) no.11, 633
doi:10.1140/epjc/s10052-016-4493-y
[arXiv:1610.02272 [hep-ph]].

\bibitem{Flett:2019pux}
C.~A.~Flett, S.~P.~Jones, A.~D.~Martin, M.~G.~Ryskin and T.~Teubner,
Phys. Rev. D \textbf{101} (2020) no.9, 094011
doi:10.1103/PhysRevD.101.094011
[arXiv:1908.08398 [hep-ph]].

\bibitem{Flett:2020duk}
C.~A.~Flett, A.~D.~Martin, M.~G.~Ryskin and T.~Teubner,
Phys. Rev. D \textbf{102} (2020), 114021
doi:10.1103/PhysRevD.102.114021
[arXiv:2006.13857 [hep-ph]].

\bibitem{Flett:2021fvo}
C.~A.~Flett, S.~P.~Jones, A.~D.~Martin, M.~G.~Ryskin and T.~Teubner,
Phys. Rev. D \textbf{105} (2022) no.3, 034008
doi:10.1103/PhysRevD.105.034008
[arXiv:2110.15575 [hep-ph]].

\bibitem{Flett:2022ues}
C.~A.~Flett, S.~P.~Jones, A.~D.~Martin, M.~G.~Ryskin and T.~Teubner,
Phys. Rev. D \textbf{106} (2022) no.7, 074021
doi:10.1103/PhysRevD.106.074021
[arXiv:2206.10161 [hep-ph]].

\bibitem{Guzey:2013xba}
V.~Guzey, E.~Kryshen, M.~Strikman and M.~Zhalov,
Phys. Lett. B \textbf{726} (2013), 290-295
doi:10.1016/j.physletb.2013.08.043
[arXiv:1305.1724 [hep-ph]].

\bibitem{Kryshen:private}
E.~Kryshen, private communication, 2016.

\bibitem{Guzey:2013qza}
V.~Guzey and M.~Zhalov,
JHEP \textbf{10} (2013), 207
doi:10.1007/JHEP10(2013)207
[arXiv:1307.4526 [hep-ph]].

\bibitem{Guzey:2020ntc}
V.~Guzey, E.~Kryshen, M.~Strikman and M.~Zhalov,
Phys. Lett. B \textbf{816} (2021), 136202
doi:10.1016/j.physletb.2021.136202
[arXiv:2008.10891 [hep-ph]].

\bibitem{Ivanov:2015hca}
D.~Y.~Ivanov, B.~Pire, L.~Szymanowski and J.~Wagner,
[arXiv:1510.06710 [hep-ph]].

\bibitem{Lappi:2020ufv}
T.~Lappi, H.~M\"antysaari and J.~Penttala,
Phys. Rev. D \textbf{102} (2020) no.5, 054020
doi:10.1103/PhysRevD.102.054020
[arXiv:2006.02830 [hep-ph]].

\bibitem{Guzey:2013jaa}
V.~Guzey, M.~Strikman and M.~Zhalov,
Eur. Phys. J. C \textbf{74} (2014) no.7, 2942
doi:10.1140/epjc/s10052-014-2942-z
[arXiv:1312.6486 [hep-ph]].

\bibitem{CMS:2022nnw}
CMS Collaboration,
CMS-PAS-HIN-22-002.

\bibitem{CMS:lumi}
  \url{https://twiki.cern.ch/twiki/bin/view/CMSPublic/LumiPublicResults}
  
\end{thebibliography}
\end{document}